
%
\def\yesans{y }
\message{ Do you have the hypertex macros, lanlmac.tex, installed on your
system (y/n)?}\read-1 to\ansm
\ifx\ansm\yesans\message{Using hypertex macros.}
\input lanlmac
\def\hrefx#1#2{\href{#1}{#2}}
\else\message{Too bad!  Using harvmac instead.}
\input harvmac
\def\hrefx#1#2{#2}
\fi

\def\hepph#1{\hrefx{http://xxx.lanl.gov/abs/hep-ph/#1}{hep-ph/#1}}
\def\heplat#1{\hrefx{http://xxx.lanl.gov/abs/hep-lat/#1}{hep-lat/#1}}

\input epsf
\noblackbox

\def\npb#1#2#3{Nucl. Phys. B #1:#2 (#3)}
\def\plb#1#2#3{Phys. Lett. B #1:#2 (#3)}
\def\prl#1#2#3{Phys. Rev. Lett. #1:#2 (#3)}
\def\pr#1#2#3{Phys. Rev. #1:#2 (#3)}
\def\prb#1#2#3{Phys. Rev. B #1:#2 (#3)}
\def\prd#1#2#3{Phys. Rev. D #1:#2 (#3)}
\def\prp#1#2#3{Phys. Rep. #1:#2 (#3)}

\def\zpc#1#2#3{Z. Phys. C #1:#2 (#3)}

\def\ap#1#2#3{Ann. Phys. (NY) #1:#2 (#3)}
\def\arnp#1#2#3{Ann. Rev. Nucl. Part. Sci #1:#2 (#3)}
\def\mpla#1#2#3{Mod. Phys. Lett. A #1:#2 (#3)}
\def\lnc#1#2#3{Lett. Nuovo Cimento #1:#2 (#3)}
\def\ijmpa#1#2#3{Int. J. Mod. Phys. A #1:#2 (#3)}

\def\app#1#2#3{Acta. Phys.  Polon. #1:#2 (#3)}
\def\pha#1#2#3{Physica #1:#2 (#3)}

\def\semi{;\item{}}
\def\etal{{\it et al.}}
\def\ibid{{\it ibid}}



\def\tr{{\rm Tr}}
\def\im{{\rm i}}

\def\tev{{\rm \, TeV}}

\def\W{{\rm\bf W}}
\def\B{{\rm\bf B}}
\def\lta{\ \hbox{\raise.55ex\hbox{$<$}} \!\!\!\!\!
\hbox{\raise-.5ex\hbox{$\sim$}}\ }
\def\gta{\ \hbox{\raise.55ex\hbox{$>$}} \!\!\!\!\!
\hbox{\raise-.5ex\hbox{$\sim$}}\ }


\catcode`\@=11

\def\slash{\global\setbox0=\hbox{\raise.075em
     \hbox{\m@th\kern-.075em $\mathchar"0236$}}%
     \wd0=0pt \ht0=0pt \dp0=0pt \box0}

\catcode`\@=12


\lref\stlimit{P. Langacker, University of Pennsylvania preprint UPR-0624T
(\hepph{9408310})\semi
see also P. Langacker and J. Erler, \item{}
\hrefx{http://www-pdg.lbl.gov/www/rpp/book/page1304.html}
{http://www-pdg.lbl.gov/www/rpp/book/page1304.html} \semi
and \prd{50}{1304}{1994}.
}

\lref\velt{M. Veltman, \app{B8}{475}{1977}.
}

\lref\lqt{B. Lee, C.  Quigg, and H. Thacker, \prl {38}{883}{1977}.
}

\lref\chan{M. S. Chanowitz, \arnp{38}{323}{1988}.
}

\lref\dpf{Report of the ``Electroweak Symmetry Breaking and Beyond the
Standard Model" working group of the DPF Long Range Planning Study, H.~Haber,
{\it et. al.}, eds.
}

\lref\thooft{ G.~'t~Hooft, in {\it Recent Developments in Gauge
Theories}, G.~'t~Hooft, {\it et. al.}, eds., Plenum Press, New York NY
1980.
}

\lref\wilson{K.~G.~Wilson, \prb {4}{3184}{1971}\semi
K.~G.~Wilson and J.~Kogut, \prp {12}{76}{1974}.
}

\lref\dn{R.~Dashen and H.~Neuberger, \prl {50}{1897}{1983}.
}

\lref\maiani{L.~Maiani, G.~Parisi and R.~Petronzio, \npb{136}{115}{1978}.
}

\lref\kuti{M.~L\"{u}scher and P.~Weisz, \npb{318}{705}{1989}\semi
J.~Kuti, L.~Lin and Y.~Shen, \prl {61}{678}{1988}\semi
A.~Hasenfratz \etal, \plb{199}{531}{1987}\semi
A.~Hasenfratz \etal, \npb{317}{81}{1989}\semi
G.~Bhanot \etal, \npb{353}{551}{1991}, 375:503 (1992) E.
}

\lref\neuberger{U.~M.~Heller, H.~Neuberger, and P.~Vranas,
\npb{399}{271}{1993}(\heplat{9207024})\semi
K.~Jansen, J.~Kuti, and C.~Liu, \plb{309}{119}{1993} (\heplat{9305003}),
\heplat{9305004}
}

\lref\higgs{For a review, see
J.~F.~Gunion, H.~E.~Haber, G.~L.~Kane and S.~Dawson,
{\it The Higgs Hunter's Guide}, Addison-Wesley, Reading, MA, 1990.
}

\lref\kominis{D. Kominis and R.~S.~Chivukula, \plb{304}{152}{1993}
(\hepph{9301222}).
}

\lref\onon{R.~S.~Chivukula, M.~J.~Dugan, and M.~Golden, \plb{336}{62}{1994}
(\hepph{9406281}).
}

\lref\rhoparameter{M.~Weinstein, \prd {8}{2511}{1973}.
}

\lref\lane{S. Weinberg, \prd{19}{1277}{1979}\semi
L. Susskind, \prd {20}{2619}{1979}\semi
For a recent review, see K.~Lane, Boston University preprint
BUHEP-94-2, in {\it 1993 TASI Lectures} (World Scientific, Singapore)
(\hepph{9401324}).
}

\lref\extc{S. Dimopoulos and L. Susskind, \npb{155}{237}{1979}\semi
E. Eichten and K. Lane, \plb{90}{125}{1980}.
}

\lref\cstdpf{R.~S.~Chivukula, E.~H.~Simmons, and J.~Terning,
report of the ``Strongly Coupled Electroweak Symmetry Breaking:
Models" subgroup of the ``Electroweak Symmetry Breaking and Beyond the
Standard Model" working group of the DPF Long Range Planning Study.
Boston University Preprint BUHEP-95-7.
}

\lref\simmons{E. Simmons, \npb{312}{252}{1989}\semi
C. Carone and E. Simmons, \npb{397}{591}{1992} (\hepph{9207273})\semi
C. Carone and H. Georgi, \prd{49}{1427}{1994} (\hepph{9308205}).
}

\lref\cdg{R.~S.~Chivukula, M. J. Dugan, and M. Golden, \prd{47}{2930}{1993}
(\hepph{9206222}).
}

\lref\topmode{Y. Nambu, Enrico Fermi Institute Preprint EFI 88-39\semi
V. A.  Miransky, M. Tanabashi, and K. Yamawaki, \plb{221}{177}{1989}
and \mpla{4}{1043}{1989}.
}

\lref\bardeen{W. A. Bardeen, C. T. Hill, and M.  Lindner,
\prd{41}{1647}{1990}.
}

\lref\fourgen{C.~T.~Hill, M.~Luty, and E.~A.~Paschos,\prd{43}{3011}{1991}\semi
T.~Elliot and S.~F.~King, \plb{283}{371}{1992}.
}

\lref\brokentc{C.~T.~Hill \etal, \prd{47}{2940}{1993} (\hepph{9201233}).
}

\lref\models{C.~T.~Hill,\plb{266}{419}{1991}\semi
S.~Martin, \prd{45}{4283}{1992} and \prd{46}{2197}{1992}\semi
N.~Evans, S.~King, and D.~Ross, \zpc{60}{509}{1993}.
}

\lref\setc{T.~Appelquist \etal, \plb{220}{223}{1989}\semi
T.~Takeuchi, \prd {40}{2697}{1989}\semi
V.~A.~Miransky and K.~Yamawaki,\mpla{4}{129}{1989}.
}

\lref\nambu{Y. Nambu and G. Jona-Lasinio, \pr{122}{345}{1961}.
}

\lref\cohen{R.~S.~Chivukula, A.~Cohen, and K.~Lane, \npb{343}{554}{1990}.
}

\lref\gk{D. B. Kaplan and H. Georgi, \plb{136}{183}{1984}\semi
D. B. Kaplan, S. Dimopoulos and H. Georgi, \plb{136}{187}{1984}\semi
H. Georgi, D. B. Kaplan and P. Galison \plb{143}{152}{1984}\semi
H. Georgi and D. B. Kaplan, \plb{145}{216}{1984}\semi
M. J. Dugan, H. Georgi and D. B. Kaplan, \npb{254}{299}{1985}.
}

\lref\vk{R.~S.~Chivukula and V.~Koulovassilopoulos, \plb{309}{371}{1993}
(\hepph{9304293}) and \prd{50}{3218}{1994} (\hepph{9312317}).
}


\lref\wein{S. Weinberg, \pha{96A}{327}{1979}\semi
H. Georgi, Weak Interactions and Modern Particle Theory,
The Benjamin/Cummings Publishing Company, Inc., Menlo Park, CA, 1984\semi
A. Manohar and H. Georgi, \npb{234}{189}{1984}.
}

\lref\cgga{M.Chanowitz, M. Golden and H. Georgi, \prl{57}{2344}{1987}\semi
M.Chanowitz, M. Golden and H. Georgi, \prd{36}{1490}{1987}.
}

\lref\equiva{J. Cornwall, D. Levin and G. Tiktiopoulos, \prd{10}{1145}{1974}
\semi
C. Vayonakis, \lnc{17}{383}{1976}.
}

\lref\equivb{M. Chanowitz and M.K. Gaillard, \npb{261}{379}{1985}.
}

\lref\longh{A. Longhitano, \prd{22}{1166}{1980} and \npb{188}{118}{1981}.
}

\lref\chilg{R. Renken and M. Peskin, \npb{211}{93}{1983}\semi
M. Golden and L. Randall, \npb{361}{3}{1991}\semi
B. Holdom and J. Terning, \plb{247}{88}{1990}\semi
A. Dobado, D. Espriu, and M. Herrero, \plb{253}{161}{1991}.
}

\lref\georgi{H. Georgi, \npb{363}{301}{1991}.
}

\lref\pdg{Particle Data Group, Review of Particle Properties,
\prd{50}{1173}{1994}.
}

\lref\gasser{J. Gasser and H. Leutwyler, \ap{158}{142}{1984} and
\npb{250}{465}{1985}.
}

\lref\holdom{B. Holdom, \plb{258}{156}{1991}.
}

\lref\lps{B.W. Lynn, M.E. Peskin and R.G. Stuart, Radiative corrections in
$SU(2)\times U(1)$: LEP/SLC, {\it in} Physics at LEP, ed. J. Ellis and R.
Peccei, Report CERN 86-02, vol. 1 (CERN, Geneva, Switzerland, 1986), p.90.
}

\lref\kenlan{D. Kennedy and P. Langacker, \prl {65}{2967}{1990}\semi
W.J. Marciano and J.L. Rosner, \prl{65}{2963}{1990}\semi
G. Altarelli and  R. Barbieri, \plb{253}{161}{1991}.
}

\lref\pestak{M.E. Peskin and T. Takeuchi, \prl{65}{964}{1990}.
}

\lref\hagiwar{K. Gaemers and G. Gounaris, \zpc {1}{259}{1979}\semi
K. Hagiwara \etal, \npb{282}{253}{1987}.
}

\lref\wudka{J. Wudka, \ijmpa{9}{2301}{1994} (\hepph{9406205}).
}

\lref\lepcalcdv{S. Dawson and G. Valencia, Brookhaven Preprint BNL-60949
(1994) (\hepph{9410364}).
}

\lref\lepcalc{P. Hernandez and F.J. Vegas, \plb{307}{116}{1993}
(\hepph{9212229})\semi
K. Hagiwara \etal, KEK Preprint KEK-TH-375 (\hepph{9409380})\semi
D. Zeppenfeld, Low Energy Constraints and
Anomalous Triple Gauge Boson Couplings, Proceedings, {\it Workshop on
Physics and Experiments with Linear $e^+e^-$ Colliders}, Waikoloa,
Hawaii, April 26--30, 1993
(\hepph{9307333}).
}

\lref\egamcalc{K. Cheung \etal, \prd{51}{5}{1995} (\hepph{9403358}).
}

\lref\fnalcalc{U.Baur \etal, Electroweak Physics at Current
Accelerators and The Supercollider, Proceedings of the Workshop on
Physics at Current Accelerators and the Supercollider, Argonne, IL, 2-5
June, 1993 (\hepph{9303318})
\semi H. Aihara, $W\gamma/WZ$ Production at the Tevatron,
Proceedings of the International Symposium on Vector Boson
Self-Interactions, UCLA, Los Angeles, CA, 1-3 February, 1995
\semi T. Fuess, $WW/WZ/ZZ$ Production at the Tevatron, \ibid
\semi H. Johari, A Search for $W$ Pair Production in D0, \ibid
\semi L. Zhang, $WW\to$ Dilepton Search at CDF, \ibid\semi
D. Neuberger, First Results on $W\gamma$ Production from the Ongoing
Tevatron Run at CDF, \ibid
\semi C. Wendt, Testing Vector Boson Self-Interactions in Future Tevatron
Experiments, \ibid.
}

\lref\devi{M.B. Einhorn, Proceedings of Conference on Unified
Symmetry in the Small and in the Large, Coral Gables, FL, 25-27 Jan
1993 (\hepph{9303323})\semi
C. Arzt, M.B. Einhorn, and J. Wudka, \npb{433}{41}{1995}
(\hepph{9405214}).
}

\lref\eecalc{G. Barbellini \etal, {\it in} Physics at LEP, ed.
J. Ellis and R. Peccei, Report CERN 86-02, vol. 2 (CERN, Geneva,
Switzerland, 1986), and references therein\semi
G. Couture and S. Godfrey, \prd{50}{5607}{1994} (\hepph{9406257})\semi
F. Boudjema, Proceedings of Physics and Experiments with Linear $e^+ e^-$
Colliders, ed. by F. A. Harris et al., (1993) p. 713 and references
therein (\hepph{9308343})\semi
G. Gounaris \etal, {\it in} Proc of the Workshop on $e^+ e^-$
Collisions at 500 GeV: The Physics Potential, DESY-92-123B. p. 735,
ed. P. Zerwas\semi
M. Bilenky \etal, \npb{409}{22}{1993} and \npb{419}{240}{1994}\semi
D.L.  Burke in J. Hawthorne, Ed., Gauge Bosons and Heavy Quarks,
Procedings of the 19th SLAC Summer Institute on Particle
Physics, SLAC-REPORT-378, Stanford, 1991\semi
G.L. Kane, J. Vidal, and C.-P. Yuan, \prd{39}{2617}{1989}\semi
S. Godfrey, The Measurement of Tri-linear Gauge Boson Couplings at $e^+ e^-$
Colliders, in Proceedings, International Symposium on Vector Boson
Self-Interactions, UCLA, Los Angeles, CA, 1-3 February, 1995\semi
J. Busenitz, Measuring Anomalous Couplings at LEP II, \ibid\semi
T. Barklow, Tri-linear Couplings at the NLC, \ibid\semi
T. Barklow, SLAC-PUB-5808 (1992), published in Saariselka Workshop 1991:423
(QCD161:W579:1991)\semi
B. Holdom, \plb{258}{156}{1991}\semi
D. Espriu and M.J. Herrero, \npb{373}{117}{1992}.
}

\lref\heracalc{T.Helbig and H. Spiesberger, \npb{373}{73}{1992}\semi
C.S. Kim, J.Lee, and H.S. Song, \zpc{63}{673}{1994} (\hepph{9307377})\semi
R. Walczak, $W$ and $Z$ Boson Production at HERA, Proceedings,
International Symposium on Vector Boson Self-Interactions, UCLA,
Los Angeles, 1-3 February, 1995\semi
A. Schoning, Towards Probing the $WW\gamma$ Vertex with H1 at HERA, \ibid
}

\lref\lhccalc{G. L. Kane, J. Vidal, and C.-P. Yuan, \prd{39}{2617}{1989}\semi
J. Womersley, Probing tri-linear couplings at the LHC, Proceedings of the
International Symposium on Vector Boson Self-Interactions, UCLA, Los Angeles,
1-3 February, 1995.}

\lref\bdv{J. Bagger, S. Dawson and G. Valencia, \npb{399}{364}{1993}.
}

\lref\dvb{S. Dawson, G. Valencia, \npb{352}{27}{1991}.
}

\lref\fls{A. Falk, M. Luke,  and E. H. Simmons, \npb{365}{523}{1991}.
}

\lref\dht{A. Dobado, M.J. Herrero, J. Terron, \zpc{50}{205}{1991}.
}

\lref\bho{U. Baur, T. Han and J. Ohnemus,
Florida State University Preprint FSU-HEP-941010 (\hepph{9410266}).
}

\lref\fbj{F. Boudjema, in 2nd International Workshop on
Physics and Experiments with Linear $e^+e^-$ Colliders, Waikoloa, HI,
26-30 April, 1993 (\hepph{9308343}).
}

\lref\goldch{M.Chanowitz and M. Golden, \prl{61}{1035}{1988} and
\prl{63}{466E}{1989}.
}

\lref\golddd{J. Bagger \etal, \prd{49}{1246}{1994} (\hepph{9306256}).
}


\nfig\hidfig{
The absolute value of the $\pi\pi\to\pi\pi$ scattering
amplitude vs. CM energy for different values of $M$. Here $j = 4$, $n = 32$,
$m_\psi = 125$ GeV, and $f = 250$ GeV. The curves correspond to roughly $8M /
m_\psi \sim $ 10000, 600, 200, 100, and 60.  The curve with the leftmost bump
is 10000, and the low nearly structureless curve is $8 M / m_\psi \sim
60$. For comparison, the dashed line shows the scattering amplitude in the
limit $m_\psi \to \infty$ with $M$ adjusted to produce a Higgs resonance at
approximately 500 GeV.
}

\lref\CCWZ{S. Coleman, J. Wess, and B. Zumino, \pr{177}{2239}{1969}\semi
C. Callan \etal, \pr{177}{2246}{1969}.
}

\lref\cs{R.~N.~Cahn and M.~Suzuki, \prl {67}{169}{1991}.
}

\lref\BESS{
R.Casalbuoni \etal, \plb{155}{95}{1985} and \npb{282}{235}{1987}.
}

\lref\tehlq{E. Eichten {\it et. al.}, \prd {34}{1547}{1986}.
}

\lref\coll{R. S. Chivukula, in DPF Conf. 1988, p. 723;
R. S. Chivukula, in {\it Snowmass: DPF Summer Study 1988}, S Jensen, ed,
p. 131\semi
S. King and M. Machacek, in {\it Snowmass: DPF Summer Study 1988}, S Jensen,
ed, p. 148;
R. S. Chivukula, in {\it Johns Hopkins Workshop 1988} p. 297\semi
R. Rosenfeld, \prd{39}{971}{1989}\semi
R. Rosenfeld and J. L. Rosner, \prd{38}{1530}{1988}\semi
R. Sekhar Chivukula and Mitchell Golden, \prd{41}{2795}{1990}\semi
R. Casalbuoni, \etal, \plb{249}{130}{1991}\semi
R. Casalbuoni, \etal, \plb{253}{275}{1991}\semi
J. Bagger, T. Han, and R. Rosenfeld, in {\it Snowmass Summer Study 1990}
p. 208\semi
K. Lane and M. V. Ramana, \prd{44}{2678}{1991}\semi
R. Casalbuoni \etal, \plb{279}{397}{1992}\semi
J. Bagger \etal, \prd{49}{1246}{1994} (\hepph{9306256})\semi
R. Casalbuoni \etal, \plb{279}{397}{1992}\semi
J. Bagger \etal, \prd{49}{1246}{1994} (\hepph{9306256})\semi
R. Casalbuoni \etal, in {\it Working Group on $e^+ e^-$ Collisions
at 500-GeV: The Physics Potential, Munich, Germany, 20 Nov 1992},
\hepph{9309334}\semi
R. Rosenfeld, \mpla{9}{735}{1994} (\hepph{9310217})\semi
R. Casalbuoni, \etal, Geneva U. Preprint UGVA-DPT-1994-03-845
(\hepph{9403305})\semi
R. Rosenfeld, Northeastern preprint NUB-3086-94-TH (\hepph{9403356})\semi
M. Baillargeon, G. Belanger, and F. Boudjema, LAPP Preprint ENSLAPP-A-473-94,
(\hepph{9405359})\semi
S. King, Southampton Preprint, SHEP-93-94-27 (\hepph{9406401})\semi
M. S. Berger and M. S. Chanowitz, presented at {\it Workshop on Gamma-Gamma
Colliders, Berkeley, CA, 28-31 Mar 1994}, (\hepph{9406413})\semi
M. S. Chanowitz and W. B. Kilgore, Lawrence Berkeley Lab Preprint
LBL-36334 (\hepph{9412275}).
}

\lref\SS{M.~Soldate and R.~Sundrum, \npb{340}{1}{1990}.
}

\lref\AT{T. Appelquist, J. Terning, \prd{47}{3075}{1993} (\hepph{9211223}).
}

\lref\dugol{M. Dugan and M. Golden, \prd{48}{4375}{1993} (\hepph{9306265}).
}

\lref\reconsid{R. S. Chivukula, M. J. Dugan, M. Golden, \plb{292}{435}{1992}
(\hepph{9207249}).
}

\lref\eephenom{J. Layssac, F. M. Renard, C. Verzegnassi, \prd{49}{2143}{1994},
(\hepph{9310367}).
}

\lref\dimop{S. Dimopoulos, S. Raby, and G. Kane, \npb{182}{1981}{77}.
}

\lref\Qprob{Stephen R. Sharpe, \prd{46}{3146}{1992} (\heplat{9205020}), and
in {\it Amsterdam Lattice 1992}, p 213 (\heplat{9211005}).
}

\lref\Quenched{C. Bernard, M. Golterman, \prd{46}{853}{1992} (\heplat{9204007})
and \npb{(Proc.Suppl.) 26}{360}{1992}.
}

\lref\hidden{R. S. Chivukula and M. Golden, \plb{267}{233}{1991}.
}

\lref\canon{M. Golden, in {\it Beyond the Standard Model,
Iowa State University, Nov. 18-20, 1988}, K. Whisnant and B.-L. Young Eds.,
p. 111, World Scientific, Singapore, 1989.
}

\lref\Coleman{S. Coleman, R. Jackiw, and H. D. Politzer, \prd{10}{2491}{1974}.
}

\lref\scalars{R. S. Chivukula and M. Golden, \npb{372}{44}{1992}.
}

\lref\Einhorn{Martin B.  Einhorn, \npb {246}{75}{1984}\semi
R. Casalbuoni, D. Dominici, and R. Gatto, \plb {147}{419}{1984}.
}

\lref\gscat{J.~Bagger, S.~Dawson, and G.~Valencia, \prl{67}{2256}{1991}\semi
R.~S.~Chivukula, M.~Golden, and M.~V.~Ramana \prl{68}{2883}{1992}, 69:1291
(1992) E (\hepph{9204220}).
}

\lref\hphenom{R. Sekhar Chivukula, Mitchell Golden, and M.V. Ramana
\plb{293}{400}{1992} (\hepph{9206255})\semi
Tao Han (Fermilab), at 2nd International Workshop on Physics and Experiments
with Linear e+ e- Colliders, Waikoloa, HI, 26-30 Apr 1993
(\hepph{9307361})\semi
M.V. Ramana, (unpublished) Boston University preprint BUHEP-93-19 \item{}
(\hepph{9311351}).
}

\lref\kny{G. Kane, private communication and in
{\it Ottawa Standard Model 1992} p. 41\semi
S.~G.~Naculich and C.~P.~Yuan, \plb{293}{395}{1992} (\hepph{9206254}).
}

\Title{\vbox{\baselineskip12pt\hbox{BUHEP-95-09}\hbox{HUTP-95/A010}
\hbox{hep-ph/9503230}}}
{\vbox{\centerline{
Theory of a Strongly Interacting}
\vskip2pt\centerline{
Electroweak Symmetry Breaking Sector}}}

\centerline{R. Sekhar Chivukula$^{a,1}$,}
\centerline{Michael J. Dugan$^{a,2}$,}
\centerline{Mitchell Golden$^{b,3}$,}
\centerline{\it and}
\centerline{Elizabeth H. Simmons$^{a,4}$}
\footnote{}{$^a$Boston University, Department of Physics, 590 Commonwealth
Avenue, Boston, MA 02215}
\footnote{}{$^b$Lyman Laboratory of Physics,
Harvard University, Cambridge, MA 02138}
\footnote{}{$^1$sekhar@bu.edu}
\footnote{}{$^2$dugan@bu.edu}
\footnote{}{$^3$golden@physics.harvard.edu}
\footnote{}{$^4$simmons@bu.edu}
\vskip .4in
\centerline{\bf ABSTRACT}

In this review we discuss theories of the electroweak symmetry
breaking sector in which the $W$ and $Z$ interactions become strong at
an energy scale not larger than a few TeV.

\Date{02/95}


The standard $SU(2)_W \times U(1)_Y$ gauge theory of the electroweak
interactions is in good agreement with all current experimental data
\stlimit.  Nonetheless, there is no direct evidence that shows which mechanism
is responsible for the breakdown of this symmetry to the $U(1)$ of
electromagnetism. However, it is clear that additional clues to the physics of
symmetry breaking must appear at energies of order a TeV or lower. Consider a
thought experiment \lqt, the scattering of longitudinally polarized $W^+$ and
$W^-$:
\eqn\Io{\epsfxsize= 3.5truein \epsfbox{fig1.eps}}
Using the Feynman-rules of the electroweak gauge theory we can
calculate $W_L^+W_L^-$ scattering at tree level.  We find that this
amplitude grows like $E_{cm}^2$:
\eqn\Ii{
{\cal A} = {g^2 s \over 8 M_W^2}\left(1+\cos\theta^{*}\right)
{}~~~,
}
plus terms that do not grow with $s$.
Projecting onto the $s$-wave state, we find
\eqn\Iii{{\cal A}^{l=0} = {g^2 s \over 128\pi M_W^2}\sim
\left(\sqrt{s} \over 2.5\ {\rm TeV}\right)^2
{}~~~.
}

Unitarity implies that some new physics has to enter to cut off the
growth of this amplitude before an energy of around 2.5
TeV\lqt\velt. That is, the dynamics associated with EWSB has to appear
before that energy scale.  There are three possibilities:
\item{$\bullet$}
There may be additional particles with masses less than or of
order of a TeV, or
\item{$\bullet$}
the $W$ and $Z$ interactions may become strong at energies of order a TeV, or
\item{$\bullet$} both of the above.

This review discusses the theory of a symmetry breaking sector in
which the $W$ and $Z$ interactions become strong at or below an energy
scale of order a TeV. For an introduction to the phenomenology of a
strongly-interacting symmetry breaking sector, we refer the reader to
the review of Chanowitz \chan.  For a more detailed review of
the phenomenological situation at specific proposed colliders, such as
the LHC or NLC, we refer the reader to the sections on strongly
coupled electroweak symmetry breaking in \dpf.

In the next section we discuss theories of electroweak symmetry
breaking.  In the second section, we discuss the use of effective
Lagrangians to describe the phenomenology of a strongly-interacting
symmetry breaking sector.  In the third section, we discuss the
limitations of the effective Lagrangian framework. Our conclusions are
presented in the final section.

\newsec{Theories of Electroweak Symmetry Breaking}

Theories of electroweak symmetry breaking may be classified by the
energy scale of the dynamics responsible for the symmetry breaking.
There are theories, such as technicolor, in which the physics
responsible for symmetry breaking occurs at an energy of order a TeV,
and there are theories, such as the top mode standard model, in which
the physics is at a much higher energy.

We begin our discussion of theories of symmetry breaking with a
description of the successes and shortcomings of theories with
fundamental scalars, in particular the standard one-doublet Higgs
model. We argue that, because of triviality, any theory with
``fundamental'' scalars can only be regarded as a low-energy effective
theory for some more fundamental dynamics at a higher energy scale
which is ultimately responsible for electroweak symmetry breaking. We
further argue that when the scale of new physics is high, the
low-energy effective scalar theory is weakly-coupled and cannot give
rise to strong $W$ and $Z$ interactions at energies of order a TeV.

Next, we discuss technicolor, the prototypical theory of dynamical
electroweak symmetry breaking. In technicolor theories the scale of
the physics responsible for electroweak symmetry breaking is of order
a TeV. In contrast to theories with fundamental scalars, these
theories can give rise to strong $W$ and $Z$ interactions at energies
of order a TeV.

We conclude with a discussion of theories in which the scale of the
physics of electroweak symmetry breaking may be adjusted to a value of
order a TeV, in which case the theory is technicolor-like, or to a
much higher value, in which case the theory generally contains light
scalar particles which appear to be fundamental. As the scale of
symmetry-breaking physics is varied, the behavior of the $W$ and $Z$
scattering amplitudes interpolates between the two extremes discussed
above: when the scale of symmetry-breaking physics is of order a TeV,
the $W$ and $Z$ interactions can become strong; if the scale is
much higher they cannot.

\subsec{The Standard One-Doublet Higgs Model and Generalizations Thereof}

In the standard one-doublet Higgs model one introduces a fundamental
scalar doublet of $SU(2)_W$:
\eqn\IIai{
\phi=\left(\matrix{\phi^+ \cr \phi^0 \cr}\right)
{}~~~,
}
which has a potential of the form
\eqn\IIaii{
V(\phi)=\lambda \left(\phi^{\dagger}\phi - {v^2\over 2}\right)^2
{}~~~.
}
In the potential \IIaii, $v^2$ is assumed to be positive in order to
favor the generation of a non-zero vacuum expectation value for
$\phi$.  This vacuum expectation value breaks the electroweak
symmetry, giving mass to the $W$ and $Z$.  When symmetry breaking
takes place, the four degrees of freedom in $\phi$ divide up. Three of
them become the longitudinal components, $W_L$ and $Z_L$, of the gauge
bosons, and the fourth, commonly called $H$ (for Higgs particle), is left
over
\eqn\IIaiii{
\phi=\Omega\left(
\matrix{ 0\cr {{H+v}\over\sqrt{2}}\cr}
\right)
{}~~~.
}
In \IIaiii, $\Omega$ is an $SU(2)$ matrix.  If we make an $SU(2)_W$ gauge
transformation until $\Omega$ is the identity, we arrive at unitary gauge.

The exchange of the Higgs boson contributes to $W_L W_L$
scattering.  In the limit in which $E_{cm}$ is large compared to the
masses of the particles in the process, the leading contribution (in
energy) from Higgs boson exchange exactly cancels the bad high-energy
behavior displayed in eqn.~\Ii
\eqn\IIaxiii{{\lower15pt\hbox{\epsfxsize=2.2truein \epsfbox{fig2.eps}}}
\rightarrow
{\cal A} = -{g^2 s \over 8 M_W^2}\left(1+\cos\theta^{*}\right)
{}~~~,
}
plus terms which do not grow with energy.  At tree-level the Higgs
boson has a mass given by $m^2_H = 2\lambda v^2$.  In order for this
theory to give rise to strong $W$ and $Z$ interactions, it would be
necessary that the Higgs boson be heavy and, therefore, that $\lambda$
be large.

This explanation of electroweak symmetry breaking is unsatisfactory for a
number of reasons.  For one thing, this model does not give a dynamical
explanation of electroweak symmetry breaking: one simply assumes that the
potential is adjusted to produce the desired result.  In addition, when
embedded in theories with additional dynamics at higher energy scales,
these theories are technically unnatural \thooft\ in the following sense:
radiative effects ({\it e.g.} one-loop contributions to the Higgs mass),
are typically proportional to whatever cutoff is put on the theory
\eqn\IIaiiia{
{\lower15pt\hbox{{\epsfxsize=2.3 truein \epsfbox{fig3.eps}}}}
\rightarrow \delta m_H^2 \propto \Lambda^2
{}~~~.
}
More precisely, there is no ordinary\foot{In supersymmetric theories,
the mass of the Higgs particle(s) are protected by the chiral symmetry
of their fermionic partners. In such theories, however, the scalar
self-couplings are related to the gauge coupling constants and,
therefore, these theories do not give rise to strong $W$ and $Z$
interactions \higgs.} symmetry protecting the mass of the Higgs.  When
a fermion mass goes to zero, there is a chiral symmetry that protects
the fermion mass from getting large radiative corrections; the Higgs
mass has no such protection in the standard model. Therefore, the
parameters of the theory must be carefully adjusted in order to keep
the weak scale of order 250 GeV.  In particular, in a theory with a
higher scale, such as a Grand Unified Theory, there is no explanation
for why the Higgs mass is not equal to the GUT scale.

Perhaps most unsatisfactory, however, is that theories of fundamental
scalars are probably ``trivial'' \wilson, {\it i.e.}, it is not possible
to construct an interacting
theory of scalars in four dimensions that is valid to arbitrarily
short distance scales.  In quantum field theories, fluctuations in the
vacuum screen charge -- the vacuum acts as a dielectric
medium. Therefore there is an effective coupling constant which
depends on the energy scale ($\mu$) at which it is measured. The
variation of the coupling with scale is summarized by the
$\beta$--function of the theory
\eqn\IIaiv{
\beta(\lambda) = \mu{d\lambda\over d\mu}
{}~~~.
}
The only coupling in the Higgs sector of the standard model is the
Higgs self-coupling $\lambda$. In perturbation theory, the
$\beta$-function is calculated to be
\eqn\IIav{
{\lower15pt\hbox{\epsfysize=0.5 truein \epsfbox{fig4.eps}}}
\rightarrow \beta = {3\lambda^2 \over 2 \pi^2}
{}~~~.
}
Using this $\beta$--function and the differential equation eq.~\IIaiv,
one can compute the behavior of the coupling constant as a function of
the scale\foot{Since these expressions were computed in perturbation
theory, they are only valid when $\lambda(\mu)$ is sufficiently
small. We will return to the issue of strong coupling below.}. One
finds that the coupling at a scale $\mu$ is related to the coupling at
some higher scale $\Lambda$ by
\eqn\IIavi{
{1\over\lambda(\mu)}={1\over\lambda(\Lambda)}
+{3\over 2\pi^2}\log{\Lambda\over\mu}
{}~~~.
}
In order for the Higgs potential to be stable, $\lambda(\Lambda)$
has to be positive. This implies that
\eqn\IIavii{{1\over\lambda(\mu)} \ge {3\over
2\pi^2}\log{\Lambda\over\mu}
{}~~~.
}
Thus, we have the bound
\eqn\IIaviii{
\lambda(\mu)\le{2\pi^2\over 3\log\left({\Lambda\over\mu}\right)}
{}~~~.
}
If this theory is to make sense to arbitrarily short distances, and
hence arbitrarily high energies, we should take $\Lambda$ to $\infty$
while holding $\mu$ fixed at about 1 TeV. In this limit we see
that the bound on $\lambda$ goes to zero. In the continuum limit, this
theory is trivial; it is free field theory.

The inequality above can be translated into an upper bound on the mass
of the Higgs boson\dn. From eq.~\IIaviii\ we have
\eqn\IIaix{
{\Lambda\over\mu}\le \exp{\left(
{2\pi^2\over 3\lambda(\mu)}\right)}
{}~~~,
}
but
\eqn\IIax{
m_H^2\sim 2v^2\lambda(m_H)
{}~~~,
}
thus
\eqn\IIaxi{
\Lambda \le m_H \exp{\left({4\pi^2 v^2\over 3 m_H^2}\right)}
{}~~~.
}
For a given Higgs boson mass, there is a {\it finite} cutoff energy at which
the description of the theory as a fundamental scalar doublet stops
making sense. This means that the standard one-doublet Higgs model can
only be regarded as an {\it effective} theory valid below this cutoff.

The theory of a relatively light weakly coupled Higgs boson, can be
self-consistent to a very high energy.  For example, if the theory is
to make sense up to a typical GUT scale energy, $10^{16}$ GeV, then
the Higgs boson mass has to be less than about 170 GeV \maiani.  In
this sense, although a theory with a light Higgs boson does not really
answer any of the interesting questions ({\it e.g.}, it does not
explain {\it why} $SU(2)_W\times U(1)_Y$ breaking occurs), the theory does
manage to postpone the issue up to higher energies.

The theory of a heavy Higgs boson ({\it i.e.} with a mass of about 1
TeV), however, does not really make sense.  Since we have computed the
$\beta$-function in perturbation theory, this answer is only reliable
at energy scales at which $\lambda(\mu)$ (as well as the Higgs boson
mass) is small. Fortunately, non-perturbative lattice calculations are
available. Early estimates \kuti\ indicated that if the theory was to
make sense up to 4 TeV, the mass of the Higgs boson had to be less
than about 640 GeV.  More recent results \neuberger\ imply that this
bound may be relaxed somewhat; one might be able to get away with an
800 GeV Higgs boson, but the Higgs boson mass is certainly bounded by
a value of this order of magnitude. The triviality limits on the mass
of the Higgs boson imply that it is not possible for the $W_L$ and
$Z_L$ scattering amplitudes in the standard model to truly become
large at energies well below the cutoff.  This result is especially
interesting because it implies that if nothing shows up below energies
of the order 700--800 GeV, then something truly ``non-trivial'' is
going on. We just have to find it.

It is straightforward to generalize the one-doublet Higgs model to
models with more than one fundamental scalar doublet, or to models
with scalars in other representations of the $SU(2)_W$ \higgs. In such
theories, one or more particles with the quantum numbers of the
standard-model Higgs boson (as well as, potentially, particles of
weak-isospin 2 \higgs
\onon) contribute to $W_L W_L$ scattering. However, all such
models\foot{Supersymmetric models have a Higgs sector containing two
scalar doublets. In principle, they are trivial as well. However, as
noted above \higgs, the quartic couplings in such models are typically
quite small and the physics of symmetry-breaking may arise at much
higher scales.} suffer from the problems described above for the
one-doublet standard model. In fact, because these theories involve
more scalar degrees of freedom, they typically have $\beta$-functions
which are {\it larger} (more positive) then the standard model. For
this reason, the corresponding triviality constraints on the masses of
particles are typically {\it stronger} \kominis\onon.

In addition, in models with more than one doublet of scalars,
care must be taken to insure that the
weak-interaction $\rho$-parameter
\eqn\IIaxii{\rho={M_W\over {M_Z\cos\theta_W}}
{}~~~,
}
does not deviate significantly from one.  In the standard model, this
parameter is (at tree-level) automatically equal to one. This is the
result of an accidental symmetry \rhoparameter.  While the potential
eqn~\IIaii\ has only a manifest $SU(2)_W \times U(1)_Y$ invariance, it
is actually invariant under a global $O(4) \approx SU(2)_{L(W)}
\times SU(2)_R$ symmetry. When symmetry breaking occurs, the symmetry
breaking sector in the one-doublet Higgs model has a residual
$SU(2)_{L+R}$ ``custodial'' symmetry which ensures that the relation
$\rho=1$ is satisfied.

Finally, we note that any theory of electroweak symmetry breaking must
also allow for the symmetry breaking to be transmitted to the quarks
and leptons, so that they can become massive as well. In the standard
model, fermion masses are obtained by introducing Yukawa interactions
that couple the Higgs doublet to the left- and right-handed
fermions. After the Higgs field develops an expectation value, the
fermions obtain a mass proportional to the Yukawa coupling. By
choosing the Yukawa couplings appropriately, one can accommodate the
observed masses (and mixing angles) of the quarks and
leptons. Understanding the couplings of the fermions to the
symmetry-breaking sector, therefore, generally involves understanding
the physics of flavor symmetry breaking. As we will not be discussing
the physics of flavor here, we will have little to say about the
couplings of ordinary fermions to the symmetry-breaking sector in the
current review.

\subsec{Technicolor}

In models with fundamental scalars, electroweak symmetry breaking can
be accommodated if the parameters in the potential (which presumably
arise from additional physics at higher energies) are suitably
chosen. By contrast, technicolor theories strive to explain
electroweak symmetry breaking in terms of physics operating at an
energy scale of order a TeV.  In technicolor theories, electroweak
symmetry breaking is the result of chiral symmetry breaking in an
asymptotically-free, strongly-interacting gauge theory with massless
fermions.  Unlike theories with fundamental scalars, these theories
are technically natural: just as the scale $\Lambda_{QCD}$ arises
in QCD by dimensional transmutation, so too does the weak scale $v$ in
technicolor theories.  Accordingly, it can be
exponentially smaller than the GUT or Planck scales.  Furthermore,
asymptotically-free non-abelian gauge theories may be fully consistent
quantum field theories.

In the simplest technicolor theory one introduces a (massless)
left-handed weak-doublet of ``technifermions'', and the corresponding
right-handed weak-singlets, which transform as $N$'s of a strong
$SU(N)_{TC}$ technicolor gauge group. In analogy to the (approximate)
chiral $SU(2)_L \times SU(2)_R$ symmetry on quarks in QCD, the strong
technicolor interactions respect an $SU(2)_L \times SU(2)_R$ global
chiral symmetry on the technifermions. When the technicolor
interactions become strong, the chiral symmetry is broken to the
diagonal subgroup, $SU(2)_{L+R}$, producing three Nambu-Goldstone
bosons which become, via the Higgs mechanism, the longitudinal degrees
of freedom of the $W_L$ and $Z_L$.  Because the left-handed and
right-handed techni-fermions carry different electroweak quantum
numbers, the electroweak interactions break to electromagnetism.  If
the $f$-constant of the theory, the analog of $f_\pi$ in QCD, is
chosen to be 246 GeV, then the $W$ mass has its observed
value. Furthermore, since the symmetry structure of the theory is {\it
precisely} the same as that of the standard one-Higgs-doublet model,
the remaining $SU(2)_{L+R}$ custodial symmetry insures that, to
lowest order in the hypercharge coupling, $M_W = M_Z \cos\theta_W$.
As discussed in section 2, at low energies, the phenomenology of a
model with an $SU(2)_L \times SU(2)_R
\rightarrow SU(2)_{L+R}$ symmetry can be described in terms of an
effective chiral Lagrangian.

In addition to the ``eaten'' Nambu-Goldstone bosons, such a theory
will give rise to various resonances, the analogs of the $\rho$,
$\omega$, and possibly the $\sigma$, in QCD.  In general, the growth
of the $W_L$ and $Z_L$ scattering amplitudes (eq.~\Ii) are cut off by
exchange of these heavy resonances,
\goodbreak\medskip
\vbox{
\epsfxsize=3.5in \epsfbox{fig5.eps}
\medskip\noindent
}
\medbreak
\noindent
just as in QCD the growth of pion--pion scattering amplitudes are cut
off by QCD resonances.  Scaling from QCD, we expect that the masses of
the various resonance will be of order a TeV.  Unlike the
situation in models with only fundamental scalars in the symmetry
breaking sector, the scattering of longitudinal $W$ and $Z$ bosons can
truly be strong. In section 3 we will discuss the
resonances that can occur in these models.

The symmetry breaking sector must also couple to the ordinary
fermions, allowing them to acquire mass.  In models of a strong
electroweak symmetry breaking sector there must either be additional
flavor-dependent gauge interactions \extc, the so-called ``extended''
technicolor (ETC) interactions, or Yukawa couplings to scalars
\simmons\ which communicate the breaking of the chiral symmetry of the
technifermions to the ordinary fermions. As we are not discussing the
physics of flavor, we refer the reader to ref. \cstdpf\ for a recent
review.

The technicolor theory may possess a global chiral
symmetry group $G$ larger than $SU(2)_L \times SU(2)_R$, which breaks to a
subgroup $H$ larger than $SU(2)$.  For example, it is commonly assumed
in ETC models that the ETC interactions commute with the ordinary
strong and electroweak interactions.  In order to explain the masses
of all observed fermions these models must contain an entire family of
technifermions with standard model gauge couplings.  Such models are
referred to as one-family models and possess an approximate $SU(8)_L
\times SU(8)_R$ symmetry.  In general, all that is necessary to break
electroweak
symmetry is that the electroweak $SU(2)_W \times U(1)_Y$ gauge group is
embedded in $G$ in such a way that the only unbroken subgroup of the
electroweak interactions in $H$ is electromagnetism.

One consequence of having a larger global symmetry is that the
$f$-constant of the theory may be different from 246 GeV: if the
theory contains $N_D$ doublets, all of which contribute equally to the
$W$ and $Z$ masses, the $f$-constant must be chosen to be
$246/\sqrt{N_D}$ GeV.  Furthermore, since there are generally more
broken global symmetries than the three associated with the weak
currents, chiral symmetry breaking produces additional
(pseudo-)Nambu-Goldstone bosons. Since experiment tells us that these
extra Nambu-Goldstone bosons cannot be strictly massless, other
interactions (generally electroweak, color, or ETC) must break the
corresponding global symmetries.

Non-minimal models typically also possess a larger variety of resonances
than the one-techni-doublet model. As in the simplest technicolor
model, it is the exchange of resonances that cuts off the growth in
the $W_L$ and $Z_L$ scattering amplitudes. In theories with many
doublets (or, in general, with many flavors \cdg, see the third section),
since
the $f$-constant is generally smaller than 246 GeV, we expect that the
masses of the resonances are smaller than in the one-doublet model. In
addition, because of the existence of other pseudo-Nambu-Goldstone
bosons, there may be sizable {\it inelastic} scattering amplitudes for
$W_L$ and $Z_L$ scattering.

\subsec{Other Theories of Dynamical Electroweak Symmetry Breaking}

There are also theories in which the scale ($M$) of the dynamics
responsible for electroweak symmetry breaking can, in principle, take
any value of order a TeV or greater. We will describe two classes of
such models.

The first class of models, inspired by the Nambu--Jona-Lasinio (NJL) model
\nambu\ of chiral symmetry breaking in QCD, involve a strong, but {\it
spontaneously broken}, gauge interaction. Examples include top quark
condensate (and related) models
\topmode\bardeen\fourgen\brokentc\models, as well as models with
strong extended technicolor interactions \setc. When the strength of
the effective four-fermion interaction describing the broken gauge
interactions -- {\it i.e.} the strength of the extended technicolor
interactions in strong ETC models or the strength of other gauge
interactions in top-condensate models -- is adjusted close to the
critical value for chiral symmetry breaking, the high-energy dynamics
may play a role in electroweak symmetry breaking without driving the
electroweak scale to a value of order $M$.

The second class are the Georgi-Kaplan Composite Higgs
models \gk.  In these, all {\it four} members of a Higgs
doublet are Nambu-Goldstone bosons arising from chiral symmetry
breaking due to a strong ``hypercolor'' interaction coupling to
massless hyperfermions. In these theories $SU(2)_W \times U(1)_Y$
breaking is due to vacuum misalignment, typically because of the
presence of an extra chiral gauge interaction. By adjusting the
strength of the extra interaction responsible for the misalignment of
the vacuum, it is possible to choose the scale of chiral-symmetry
breaking of the hypercolor interactions to be larger, possibly much
larger, than 1 TeV.

The high-energy dynamics must have the appropriate properties in order
for it to play a role in electroweak symmetry breaking \cohen: If the
coupling constants of the high-energy theory are small, only
low-energy dynamics (such as technicolor) can contribute to
electroweak symmetry breaking. If the coupling constants of the
high-energy theory are large and the interactions are attractive in
the appropriate channels, chiral symmetry will be broken by the
high-energy interactions and the scale of electroweak symmetry
breaking will be of order $M$.  If the transition between these two
extremes is continuous, {\it i.e.} if the chiral symmetry breaking
phase transition is {\it second order} in the high-energy couplings,
then it is possible to adjust the high-energy parameters so that the
dynamics at scale $M$ can contribute to electroweak symmetry breaking.
The adjustment of the high-energy couplings is a reflection of the
fine-tuning required to create a hierarchy of scales.

What is crucial is that the transition be (at least approximately)
second order in the high-energy couplings.  If the transition is first
order, then as one adjusts the high-energy couplings the scale of
chiral symmetry breaking will jump discontinuously from approximately
zero at weak coupling to approximately $M$ at strong coupling.
Therefore, if the transition is first order, it will generally not be
possible to maintain any hierarchy between the scale of electroweak
symmetry breaking and the scale of the high-energy dynamics.

If the transition is second order and if there is a large hierarchy of
scales ($M \gg$ 1 TeV), then close to the transition the theory
may be described in terms of a low-energy effective Lagrangian with
composite ``Higgs'' scalars -- the Ginsburg-Landau theory of the
chiral phase transition.  However, if there is a large hierarchy, the
arguments of triviality given in the first section apply to the
effective low-energy Ginsburg-Landau theory describing the composite
scalars: the effective low-energy theory would be one which describes
a {\it weakly} coupled theory of (almost) fundamental scalars, despite
the fact that the ``fundamental'' interactions are
strongly self-coupled!

For this reason, only models in which $M$ is of order 1 TeV can
result in strong $W_L$ and $Z_L$ scattering amplitudes. In these
models, while the extra ``Higgs'' scalars may be relatively heavy,
they may still be light enough that they should be included in an
effective-Lagrangian description of low-energy $W_L$ and $Z_L$
interactions. Furthermore, the interactions of these scalars can
differ significantly from those of the standard-model Higgs boson \vk.
The effective Lagrangian appropriate for describing the
phenomenology of these models is discussed in section 3.


\newsec{Effective Lagrangians and Electroweak Symmetry Breaking}

The unknown high-energy physics responsible for electroweak symmetry
breaking both provides the weak bosons with mass and influences their
interactions with one another and with other particles.  Hence, a
meticulous investigation of the properties of the weak bosons can
provide clues to the nature of the symmetry breaking sector.  The most
efficient way of proceeding is to identify a model-independent method of
analyzing the relationship between the weak bosons' properties and the
high-energy physics responsible for electroweak symmetry breaking.  We
discuss here the formalism of effective Lagrangians which will enable us
to focus on the known symmetry properties of the broken theory and to
classify interactions at energies below the symmetry-breaking scale in
terms of their transformation properties under the symmetry remaining at
low energies.  This emphasis on symmetry will enable us to make
quantitative statements about strongly-interacting dynamics for which
direct calculation is problematic.

\subsec{Effective Lagrangians}

An ``effective'' Lagrangian is one that affords an approximate
description of physics at energies below a designated cutoff scale
$\Lambda$.  The particle content and symmetry structure of the effective
Lagrangian are dictated by what exists at scales below the cutoff.  The
presence of higher-energy physics and heavier particles is incorporated
via the inclusion of appropriate non-renormalizable terms.  The terms in
an effective Lagrangian are arranged as an expansion in powers of
momentum over the cutoff, $\Lambda$. Although there are an infinite
number of terms in this expansion, at low energy the first few terms can
give a good approximation.  A familiar example of an effective
Lagrangian is the $V-A$ description of the charged-current weak
interactions at energies below $M_W$.  The effective theory includes
non-renormalizable four-fermion contact interactions that result from
``integrating out'' the propagating $W$ boson that is present at higher
energies.

The effective Lagrangian is in general non-renormalizable.  That means
that if calculated to an arbitrary number of loops, the renormalization
of the theory would require an infinite number of counterterms.  There
must be some organizational principle by which some of the operators are
included and others neglected in a particular calculation.  Moreover
this procedure has to be systematic, so that large contributions are not
neglected at any order in the expansion.

In general, the requirement is that $\Lambda$ be much larger than the
momentum scale $p$ at which the experiments are performed, and
amplitudes are written as a power series in $p/\Lambda$.  When one
computes a low-energy amplitude to a given accuracy, we compute the
required numbers of terms in the momentum expansion.  For example, in
the $V-A$ theory, the cutoff scale is the mass of the $W$, and the
momentum expansion is in terms of four-fermi operators that contain
extra derivatives and are suppressed by additional powers of $1/M_W^2$.
This expansion can be expected to work for momenta up to the cutoff.

In addition to powers of $\Lambda$ suppressing the higher dimension
operators, each operator has a dimensionless coefficient $C$.  But if an
operator in this expansion had a coefficient $C$ very much greater than
order 1, there would be some momentum scale $p \ll \Lambda$ at which
it could compete with lower dimension operators.  This would imply
that the momentum expansion had broken down at $p$, well below
$\Lambda$.  Accordingly, every dimensionless coefficient in the
expansion is expected to be smaller than or of order one, at least if
the cutoff is really $\Lambda$.

It is possible to judge whether a given experiment can place useful
limits on the coefficients of terms in the effective Lagrangian.  Say,
for example, that given the cutoff $\Lambda$ we expect a particular
coefficient $C$ to be of order 1. If a proposed experiment can only
place an upper bound of 100 on that coefficient, the measurement is
not likely to be informative.  On the other hand, if an experiment
appeared to measure a definite value of 50 for that coefficient, it
would indicate that new physics enters at a scale lower than expected
-- an informative outcome indeed!

As mentioned above, the longitudinal modes of the $W$ and $Z$ are the
Nambu-Goldstone bosons of a spontaneously broken $SU(2)\times U(1)$
symmetry.  As we will see, their lowest-order interactions are
completely determined by the symmetry structure.  Therefore,
distinguishing among different models of symmetry breaking will
require more precise measurements than might seem necessary at first
glance, because any dynamics that {\it is} sensitive to the precise
nature of the symmetry-breaking sector is suppressed by powers of
$1/\Lambda^2$.  As in the $V-A$ example, the cutoff is at the mass of
the physics that was integrated out -- a characteristic scale of the
symmetry breaking.  For example, in the standard model with $M_W\ll
m_H\lta 1\tev$, $\Lambda=m_H$, while in a technicolor model, $\Lambda$
might be of order the mass of the lightest techni-resonance.

Interestingly, there are additional constraints on the effective
Lagrangian: the cutoff scale $\Lambda$ may not get arbitrarily large,
and the $C$'s of the operators cannot get too small.  This is once again
due to the non-renormalizable nature of the theory.  The operators that
appear at any given order in the momentum expansion are needed as
counterterms for loop diagrams involving lower-order operators.  If the
cutoff $\Lambda$ were very large or a particular $C$ were very small, it
would imply that the corresponding higher-dimension operator was
unimportant.  On the other hand, the operator is a counterterm for loop
diagrams involving lower-dimension operators, and so it is unnatural to
assume that the small renormalized value of the coefficient is the
result of a cancellation of a large bare coupling with a large loop
diagram.  It is more natural to assume that the coefficients in the
effective Lagrangian are not too small, and $\Lambda$ is not too
big. This argument, known as Naive Dimensional Analysis (NDA), implies
that, for electroweak symmetry breaking, $\Lambda \lta 4\pi v$.  If this
limit is saturated, the coefficients $C$ are of order one\wein.

Our discussion in this section of the effective Lagrangian for
electroweak symmetry breaking is subject to the following constraints.
We will assume that the longitudinal $W$ and $Z$ are the only quanta in
the strongly-interacting symmetry-breaking sector that are light
compared to the symmetry-breaking scale.  This necessarily constrains
the global symmetry-breaking pattern to be $SU(2)_L\times SU(2)_R /
SU(2)_{L+R}$ or $SU(2)_L\times U(1)_R / U(1)_{L+R}$ \cgga .  We remain
mindful that the formalism is only valid in the energy regime in which
the momentum expansion is valid.

In discussing high-energy tests of the strongly-interacting
symmetry-breaking sector, we shall also rely on the ``equivalence
theorem'' \lqt\equiva\equivb .  This states that in calculating scattering
amplitudes at center-of-mass energy $E$, one may replace external
longitudinal $W$ and $Z$ bosons by the corresponding Nambu-Goldstone
bosons, up to corrections of order $M_W/E$.  The resulting
simplification is of particular use in discussing the two-body
scattering of longitudinal weak bosons.

\subsec{The effective Lagrangian at order $p^2$}

Our next task is to construct an effective Lagrangian that will enable
us to study the interactions of the $W$ and $Z$ bosons.  We consider the
most general Lagrangian consistent with the observed symmetry breaking
pattern.  We begin by considering a Lagrangian for global symmetry
breaking, in terms of the ``eaten'' Nambu--Goldstone bosons $\pi^a$.
These fields are most conveniently written in the non-linear
representation
\eqn\siga{
\Sigma = \exp(2 i \pi^a T^a / f)
{}~~~.
}
Here the $T^a$ are $SU(2)$ generators normalized to $\tr\left[ T^a
T^b\right] = \delta^{ab}/2$, and $f$ is the analogue of the pion decay
constant.  Under a global chiral transformation,
the field $\Sigma$ transforms as $\Sigma \to L \Sigma R^\dagger$, with
$L \in SU(2)_L$ and $R \in SU(2)_R$ or $U(1)_R$.

If the only low-energy degrees of freedom of interest are the
Nambu-Goldstone boson fields themselves, the most general chirally
invariant Lagrangian can be written as an expansion in powers of
derivatives of $\Sigma$ \longh\chilg .  There are no nontrivial chirally
invariant terms involving no derivatives.  And there are only two terms
with two derivatives\cgga:
\eqn\lochiralun{
{\cal L}^u_2 = {f^2 \over 4}
\tr\ \partial^\mu\Sigma^\dagger\partial_\mu \Sigma
+  {f^2 \over 2} (\rho - 1)
\left[ \tr T_3 \Sigma^\dagger \partial^\mu \Sigma \right]^2
{}~~~.
}
Here $\rho$ is an arbitrary coefficient; we will see below
that it corresponds precisely to the $\rho$ parameter defined in
eqn.~\IIaxii\ above.  Note that the second term is only invariant under
$U(1)_R$ and not the full $SU(2)_R$ symmetry group.  Terms with
more derivatives are suppressed by inverse powers of the momentum
cutoff corresponding to the scale, $\Lambda$, at which additional
physics enters; we will discuss these operators shortly.

Up to now the discussion has been about global symmetries only, but to
study the interactions of the weak bosons, one gauges the chiral
symmetries, identifying $SU(2)_L$ with $SU(2)_W$ and the diagonal
generator of $SU(2)_R$ (or the generator of $U(1)_R$) with $U(1)_Y$, and
employs the corresponding gauged Lagrangian.  To lowest order
this amounts to gauging \lochiralun,
\eqn\lowchir{
   {\cal L}_2 = {f^2 \over4} \tr\left[D^\mu
   \Sigma^{\dagger}D_\mu \Sigma\right] +  {f^2 \over 2} (\rho - 1)
\left[ \tr T_3 \Sigma^\dagger D^\mu \Sigma \right]^2
{}~~~,
}
where the
covariant derivative is $D_\mu \Sigma=\partial_\mu \Sigma+\im g\W_\mu
\Sigma-\im \Sigma g'\B_\mu$. and the gauge boson fields are
$\W_\mu=W_\mu^a T^a$ and $\B_\mu=B_\mu T^3$.
The full effective Lagrangian for the theory of gauge and Nambu-Goldstone
bosons is the sum of the lowest-order Lagrangian \lowchir, the
usual gauge-boson kinetic energy terms
\eqn\lowgauge{
\eqalign{
{\cal L}_{gauge}
= & -{1\over2}\tr\left[\W^{\mu\nu}\W_{\mu\nu}\right]-
 {1\over2}\tr\left[\B^{\mu\nu} \B_{\mu\nu}\right],\cr
&\B_{\mu \nu}\ = \ \left(\partial_\mu \B_\nu
-\partial_\nu \B_\mu \right) T^3 \cr
&\W_{\mu \nu}\ = \ \left(\partial_\mu \W_\nu -
\partial_\nu \W_\mu
-{i\over 2}\,g [\W_\mu,\W_\nu]\right)
{}~~~,
}
}
and gauge-fixing and Fadeev-Popov ghost terms.

To find expressions for the gauge boson masses in this
effective theory, we rewrite the order $p^2$ Lagrangian in
unitary gauge (where $\Sigma = 1$) and diagonalize the $W_3 - B$ mixing
matrix.  The result is
\eqn\massmat{
{g^2 f^2\over 4} W_-^\mu W_{+\mu} + {g^2 f^2\over{8 \rho \cos^2\theta}}
Z^\mu Z_\mu
{}~~~.
}
The photon $A_\mu = \sin\theta W^3_\mu + \cos\theta B_\mu$ is massless.
Since the mass of the $W$ boson is $M_W = g f/2$, the Nambu-Goldstone
boson decay constant $f$ is equal to $v \equiv 246$ GeV.  As noted
earlier, the parameter $\rho$ equals 1 for a theory in which a custodial
symmetry $SU(2)_{L+R}$ remains after chiral symmetry breaking;
otherwise the deviation of $\rho$ from 1 measures the degree of custodial
symmetry violation in the theory.

We can also obtain definitive expressions for two-body
scattering of the Nambu-Goldstone bosons ($W_L$ and $Z_L$) that, like
the $W$ and $Z$ masses, depend on $v$ and $\rho$.  In the energy range
where the effective Lagrangian and the equivalence theorem are
both valid, $M_W^2 << s << \Lambda^2$, this can be done by
expanding the Lagrangian \lochiralun\ to determine the 3-$\pi$ and
4-$\pi$ vertices, and then forming the amplitudes.  The result is
\eqn\lowethr{
\eqalign {
{\cal M}[W^+_L W^-_L \to W^+_L W^-_L] =& {i u \over {v^2 \rho}} \cr
{\cal M}[W^+_L W^-_L \to Z_L Z_L] =& {i s \over {v^2}} \left( 4 -
{3\over\rho}\right)\cr
{\cal M}[Z_L Z_L \to Z_L Z_L] =& 0
{}~~~.
}
}
and the expressions for the $W^\pm_L Z_L$ and $W^\pm_L W^\pm_L$ channels
follow by crossing symmetry.  What is striking is that these tree-level
expressions for longitudinal gauge boson scattering at energies below the
symmetry-breaking scale will be identical for {\it any} theory with an
$SU(2)_L\times SU(2)_R$ global symmetry structure at high energies.  Hence,
the expressions \lowethr\ are known as the ``low-energy theorems'' for a
strongly interacting symmetry-breaking sector.

A wealth of data from LEP, SLAC and Fermilab now tell us that $\rho$
equals 1 to a few parts in a thousand \stlimit\pdg\ :
\eqn\rhoval{
\rho - 1 = \pm .004
{}~~~.
}
Therefore, for the rest of this article we shall assume that the pattern of
symmetry breaking is $SU(2)_L\times SU(2)_R/ SU(2)_{L+R}$; the
custodial symmetry that enforces $\rho = 1$ is present.  The only
source of custodial symmetry breaking in our effective Lagrangian will
be the non-zero hypercharge coupling, $g'$.

\subsec{The effective Lagrangian at order $p^4$}

So far, we have constructed an effective Lagrangian whose
predictions depend only on the symmetries of the electroweak symmetry
breaking sector.  In order to probe other properties and differentiate
among competing models, it will be necessary to include terms in the
Lagrangian that arise at higher order in the momentum expansion.

The next-to-leading order effective Lagrangian for the Nambu-Goldstone
fields includes several terms containing four derivatives \longh\gasser :
\eqn\lagfour{
{\cal L}^u_4 = {L_1\over 16\pi^2}
 \{\tr(\partial^\mu\Sigma^\dagger\partial_\mu\Sigma)\}^2 +
 {L_2\over 16\pi^2}
 \{\tr(\partial^\mu\Sigma^\dagger\partial^\nu\Sigma)\}^2
{}~~~.
}
All other possible four-derivative terms are linear combinations of these
two or vanish by the equations of motion.

The coefficients $L_1$ and $L_2$ are new parameters of the effective
Lagrangian which are not determined by the low-energy terms. The
coefficients $L_i/16\pi^2$ of the operators in \lagfour\ are of order
$v^2/\Lambda^2$.  Therefore, the $L_i$ are of order one in a theory in
which $\Lambda\approx 4\pi v$.  NDA implies that $\Lambda$ cannot be
larger than this value.  Different underlying theories of the
high-energy physics responsible for electroweak symmetry breaking will
predict different values for the $L_i$.  It is by measuring the
physical observables related to these coefficients that experiments will
be able to constrain such models. If the $L_i$ are found to be
significantly larger than one, the scale $\Lambda$ is less than $4\pi
v$.

Again, if we are interested in studying the loop-level properties of
scattering amplitudes involving the weak bosons, we employ a gauged
effective Lagrangian. This
looks like:
\eqn\chiral{
\eqalign{
{\cal L}_4\ =&\  {L_1 \over 16 \pi^2}\, \left[
\tr {D^\mu \Sigma^\dagger D_\mu \Sigma} \right]^2
\ +\  {L_2 \over 16 \pi^2}\, \tr {D_\mu \Sigma^\dagger D_\nu \Sigma}
\tr {D^\mu \Sigma^\dagger D^\nu \Sigma} \cr
\ & \ - i g {L_{9L} \over 16 \pi^2}\, \tr {\W^{\mu \nu} D_\mu
\Sigma D_\nu \Sigma^\dagger}
\ -\ i g' {L_{9R} \over 16 \pi^2}\, \tr {\B^{\mu \nu}
D_\mu \Sigma^\dagger D_\nu\Sigma} \cr
\ & +\ g g' {L_{10}\over 16 \pi^2}\, \tr {\Sigma \B^{\mu \nu}
\Sigma^\dagger \W_{\mu \nu}}
{}~~~.
}
}
Unlike \holdom , we are not restricting ourselves to vectorial
models with $L_9^L = L_9^R$.

We now relate these various coefficients to physical quantities that
colliders are currently measuring or hope to bound in the future.  We
shall address sequentially the information provided by 2-point,
3-point and 4-point vertices involving gauge and Nambu-Goldstone
bosons.

\medskip
\noindent{$\bullet${ \tt 2-point vertices}}
\medskip

Radiative corrections from non-standard physics that alter the vacuum
polarization of the electroweak gauge bosons are known as ``oblique''
corrections \lps .  Due to their effects on many
well-measured quantities, the oblique corrections provide some of the
most important limits on the electroweak symmetry breaking
sector \chilg\kenlan\pestak .

It is conventional to describe the oblique corrections in terms of
three ultraviolet-finite combinations of vacuum polarizations \pestak:
\eqn\stuuu{\eqalign{
 \alpha S \equiv& 4 e^2 \left[ \Pi'_{33}(0) -
\Pi'_{3Q}(0)\right]\cr
\alpha T \equiv& {g^2 \over {\cos^2\theta m_Z^2}} \left[\Pi_{11}(0) -
\Pi_{33}(0)\right] \cr
\alpha U \equiv& 4 e^2 \left[\Pi'_{11}(0) -
\Pi'_{33}(0)\right]
{}~~~.
}
}
where $\Pi'(q^2) \equiv {d \Pi(q^2) / d q^2}$. After calculating radiative
corrections to an observable $x$, one can write
\eqn\stuuc{
x = x_{sm}(m_t, m_H) + \lambda^x_1 S + \lambda^x_2 T +
\lambda^x_3 U
{}~~~,
}
where $x_{sm}$ includes all standard model contributions to $x$ for
given masses of the top quark and Higgs boson, and the $\lambda^x_i$
are coefficients independent of $m_t$ and $m_H$.  When the observables
$\alpha$, $G_F$ and $M_Z$ are used to define the parameters $g$,
$g'$ and $v$ in the electroweak theory, $\alpha^x_3$ is zero for
all neutral-current and low-energy observables.  The only measured
quantity depending on $U$ is the ratio of the $W$ and $Z$
masses \pestak; furthermore, since we are assuming an
approximate custodial symmetry holds, $U/T \sim m_Z^2/\Lambda^2 << 1$.
If one takes $U \approx 0$, the $S$ parameter measures
weak-isospin-conserving oblique corrections from new physics and $T$
measures weak-isospin-violating contributions.

Examining the effective Lagrangian \chiral\ in unitary gauge, we find
that the only term that includes a 2-point vertex is the operator with
coefficient $L_{10}$.  This, then, is the only operator that
contributes to the oblique corrections at order $p^4$.  Since the
$L_{10}$ term contributes an amount $-q^2
L_{10}(M_Z)/16\pi^2$ to the vacuum polarization $\Pi_{33} - \Pi_{3Q}$,
one has \georgi
\eqn\sparamt{
L_{10} = - {1\over\pi} S
{}~~~.
}
We will find that this correspondence between $L_{10}$ and $S$
means that $L_{10}$ is better constrained at present than any of the other
$L_i$.

The $T$ parameter as defined above is related to the isospin-violating
parameter $\rho$ encountered in the discussion of weak gauge boson masses
\eqn\teero{
\alpha T = \rho - 1
{}~~~.
}
We have already limited our discussion to theories in which the
presence of an approximate custodial $SU(2)_{L+R}$ symmetry enforces
$\rho\approx 1$.  A non-zero value for the hypercharge coupling does
break the custodial symmetry, so that loop diagrams involving exchange of
hypercharge bosons do contribute to non-zero $T$.  If one is studying
the energy range $M_Z < E < m_t$, where the top quark is not present
in the effective theory, then the absence of a partner for the bottom
quark introduces additional contributions to $T$.

Current limits on S and T derived from a global fit to data
\stlimit\ are
\eqn\sval{
S = -0.15 \pm 0.25^{-0.08}_{+0.17}, \ \ \ \
T = -0.08 \pm 0.32^{+0.18}_{-0.11}
{}~~~.
}
This implies the constraint
\eqn\sval{
-.09 < L_{10} < 0.15
}
on the effective Lagrangian at order $p^4$.

\medskip
\noindent{$\bullet${\tt 3-point vertices}}
\medskip

A popular topic in recent years \foot{see e.g. Proceedings,
International Symposium on Vector Boson Self Interactions, ed. by
U. Baur, S. Errede, T. Muller, UCLA, Feb. 1-3, 1995.} has been the
study of the ability of collider experiments to test the form and
strength of the three-weak-gauge-boson vertices.  While much effort
has been devoted to studying the potential of FNAL, LEP, LEP II, LHC,
HERA and various NLCs for measuring small deviations from the
standard model predictions.  It seems clear that the prospects are dim
\devi.  Simply put, the only values of the $L_i$ that would be
accessible to {\it any} current experiment are so large that for any
reasonable $\Lambda$ they contradict the rules discussed in section 2.1,
which are an intrinsic part of the effective Lagrangian.  A similar
statement can be made for any but the highest energy experiments being
planned.  If any experiment at FNAL, LEP, LEP II, or HERA were to
measure a deviation from the standard model predictions, it
would imply an $L_i$ so large that the scale of new physics would have
to be nearly as small as $M_W$, invalidating the entire effective
Lagrangian approach.

In order to study non-standard contributions to the three-gauge-boson
vertices, we expand the effective Lagrangian \chiral\ in unitary gauge
and extract the terms with three-point vertices.  To make contact with
the literature on this topic, it is convenient to organize the
three-point terms as follows:
\eqn\lagran{
\eqalign{
{\cal L}^{3-point}_4 =
   &-\im e_*{\cos\theta\over \sin\theta}g_Z\left( W^\dagger_{\mu\nu}W^\mu
    -W_{\mu\nu}W^{\mu\dagger}\right) Z^\nu\cr
   &-\im e_*\left( W^\dagger_{\mu\nu}W^\mu-W_{\mu\nu}W^{\mu\dagger}\right)
    A^\nu\cr
   &-\im e_*{\cos\theta\over \sin\theta}k_ZW^\dagger_\mu W_\nu Z^{\mu\nu}
    -\im e_*k_\gamma W^\dagger_\mu W_\nu A^{\mu\nu}
{}~~~,
}
}
where
\eqn\params{
\eqalign{
   g_Z&={e_*^2L_9^L\over32\pi^2\sin^2\theta\cos^2\theta}+{e_*^2L_{10}\over
        16\pi^2\cos^2\theta(\cos^2\theta-\sin^2\theta)},\cr
   k_Z&={e_*^2(\cos^2\theta L_9^L-\sin^2\theta
L_9^R)\over16\pi^2\cos^2\theta\sin^2\theta}
        +{2e_*^2L_{10}\over16\pi^2\cos^2\theta\sin^2\theta},\cr
   k_\gamma&= -{e_*^2\left(
        L_9^L+L_9^R+2L_{10}\right)\over32\pi^2\sin^2\theta}
{}~~~.
}
}
The coupling $e_*$ and mixing angle are defined by
\eqn\hola{\eqalign{
e_*^2/4\pi = &\alpha_*(M_Z)\cr
\sin^2\theta \cos^2\theta \equiv & {\pi \alpha_* \over{{\sqrt2} G_F m_Z^2}}
{}~~~.
}
}

Before discussing possible experimental limits, we should demonstrate
the relationship between our effective Lagrangian and a related
formalism often used for discussion of weak boson three-point
vertices.  The notation introduced in \hagiwar\  for describing
non-standard C and P conserving contributions to weak-gauge-boson
self-interactions is
\eqn\hagiwara{
\eqalign{
{i\over e \cot\theta} {\cal L}_{WWZ} = & g_1
(W^\dagger_{\mu\nu} W^\mu Z^\nu - W^\dagger_\mu Z_\nu W^{\mu\nu}) + \kappa_Z
W^\dagger_\mu W_\nu Z^\mu\nu + {\lambda_Z\over
M_W^2}W^\dagger_{\lambda\mu}W^\mu_\nu Z^{\nu\lambda}\cr
{i\over e} {\cal L_{WW\gamma}} = & (W^\dagger_{\mu\nu} W^\mu A^\nu -
W_\mu^\dagger A_\nu W^{\mu\nu}) + \kappa_\gamma W^\dagger_\mu W_\nu
F^\mu\nu + {\lambda_\gamma\over M_W^2} W^\dagger_{\lambda\mu}W^\mu_\nu
F^{\nu\lambda}
{}~~~.
}
}
In the standard model, one has $g_1 = \kappa_Z = \kappa_\gamma= 1$ and
$\lambda_Z = \lambda_\gamma = 0$; deviations from these values are
intended to parametrize the contributions of new physics.
By comparing \hagiwara\ with \chiral\ and \params\ above, we find that
$g_1$, $\kappa_Z$ and $\kappa_\gamma$ are related to the $L_i$ by
\eqn\relat{
\left.\matrix{g_1 - 1\cr \kappa_Z - 1\cr \kappa_\gamma - 1}\right\}
\ \approx\ {\alpha_* L_i \over 4\pi \sin^2\theta}
{}~~~.
}

If the $L_i$ are of order 1, the parameters $\kappa_\gamma$ and
$\kappa_Z$ differ from unity by an amount that is of order $10^{-3}$.
It will be {\it crucial} to bear this in mind when evaluating
experimental measurements of deviations of three-point vertices from
the standard model predictions.

The coefficients $\lambda_\gamma$ and $\lambda_Z$ in \hagiwara\
accompany terms that are of higher order, $p^6$, in the momentum
expansion. Therefore, they are related not to the $L_i$ discussed above,
but to coefficients of higher-order operators.  Because we are
constructing our effective Lagrangian \chiral\ as a systematic expansion
in powers of $p^2/\Lambda^2$, if we were to include terms of order
$p^6$, they would naturally be suppressed by a factor of $1/\Lambda^2$.
For example, our order-$p^6$ Lagrangian would include a term like:

\eqn\lsixterm{
{C v^2 \over \Lambda^4}\,
\tr\left[ D_\mu \Sigma^\dagger D^\mu \Sigma\right]^3
{}~~~,
}
where $C$ is order 1.  Expressing the order $p^6$ terms in \hagiwara\
as part of such an effective Lagrangian, we have
\eqn\comparr{
{\lambda_{Z,\gamma} \over M_W^2} = {C v^2\over\Lambda^4}
{}~~~,
}
This is consistent with the fact that we expect the effects of a
strongly-interacting symmetry breaking sector to vanish in both the
limit of vanishing $W$ mass (since no symmetry-breaking will have been
effected) and in the limit of a large symmetry-breaking scale.
For $\Lambda \sim 1$ TeV, we expect
\eqn\comparrd{
\lambda_{Z,\gamma} = C {M_W^2\over\Lambda^4} \approx 10^{-4}
{}~~~.
}
Again, the small value expected for $\lambda_{Z,\gamma}$ will strongly
influence our assessment of the utility of planned experimental
measurements.

Much has been written about how to use present or anticipated data to
constrain 3-gauge-boson vertices; a compendium of results from
energies high and low appears in \wudka\ .  We shall summarize
the salient points and indicate where the interested reader may look
for further details.  We have chosen this route in large part because
most present and anticipated limits on the three-point $L_i$ (or
equivalently on the $\lambda_i$ and $\kappa_i$) are woefully loose.

A straightforward calculation starting from the effective Lagrangian
\lagran\ reveals the contribution that higher-dimension operators make
to scattering processes involving three-vector-boson vertices.  It has
been demonstrated that the various operators make complementary
contributions to different processes.  Production (at an $e^+e^-$ or
hadron collider) of pairs of $Z$ bosons or of a $W^\pm W^\pm$ final
state does not involve a three-gauge-boson vertex, and so is
independent of the $L_i$ considered here.  The process $f\bar f \to W^\pm Z$
involves $L_{9L}$; the channel $f\bar f \to W^\pm W^\mp$ involves
$L_{9L}$ and $L_{9R}$; the channel $f\bar f \to W^\pm \gamma$ involves
$L_{9L}, L_{9R}$ and $L_{10}$ \fnalcalc\eecalc\lhccalc\bdv\dvb\fls.
Precision  measurements
of $Z$ decays at LEP are indirectly sensitive to $L_{9L}$, $L_{9R}$
and $L_{10}$ \lepcalcdv\lepcalc ; measurements of $e p
\to  \nu\gamma X$ at HERA
could also potentially access those three $L_i$ \heracalc.  At an $e\gamma$
collider, the process $e\gamma \to \nu W Z$ is affected by the
$L_{9L}$ and $L_{9R}$ couplings \egamcalc.

When the cross-sections are compared with existing or projected data,
the following pattern emerges. The integrated luminosity
accumulated at the Tevatron should restrict the
$\vert\kappa_i - 1\vert$ and $\vert\lambda_i\vert$ to be smaller than
about 1 \fnalcalc\bho .  The limits from HERA are, perhaps, a little
looser at present \heracalc. In other words, the constraints derivable from
existing data greatly exceed the natural values of the
coefficients\foot{Current LEP data have been shown to place indirect bounds
of order 20 on $\vert L_{9L}\vert$ and of order 80 on $\vert
L_{9R}\vert$ \lepcalcdv\lepcalc.  These bounds are based on loop-level
calculations assuming that no large tree-level contribution causes a
significant cancellation of the effect.}

This situation will gradually improve at future colliders \devi .
Experiments at LEPII may improve the bounds on $\vert\kappa_i -1\vert$
and $\vert\lambda_i\vert$ to something of order $0.1$ \eecalc, which
would imply new strongly coupled physics at $\Lambda \approx 300$ GeV.
Either the LHC \bdv\dvb\fls\lhccalc\bho\ or an NLC \eecalc\ with a
center-of-mass energy of half a TeV could push this to roughly $0.01$,
which would imply strongly interacting new physics at $\Lambda \approx
1$ TeV.  It would take an NLC with $\sqrt{s} \ge 1$ TeV to probe
$\vert\kappa_i - 1\vert$ or $\vert\lambda_i \vert$ to anything near
their minimum size of a few times $10^{-3}$ \eecalc .  In other words,
only the highest-energy electron-positron colliders being discussed
today would have the resolution required to probe $\Lambda \approx 4\pi
v$.

\medskip
\noindent{$\bullet$ {\tt 4-point vertices}}
\medskip

Direct tests of the four-point vertices must await the
advent of high-energy colliders
capable of producing large numbers of high-momentum weak boson pairs.
Two-body scattering of weak bosons occurs at high-energy
colliders like the LHC or NLC when gauge bosons are
radiated from the incoming fermions and then rescatter via a
four-point vertex.
The four-point
vertices of greatest interest for experimentally probing the nature of
the electroweak symmetry-breaking sector are those involving only
longitudinal gauge bosons. The $V_L V_L \to V_L V_L$ processes that
they mediate are precisely those which the dynamics associated with
electroweak symmetry breaking must unitarize at an energy of 2.5 TeV
or less.  As can be seen by inspecting the effective Lagrangian
\chiral, four-point vertices involving transverse, as well as
longitudinal, gauge bosons will be affected by the higher-order terms.
However, scattering processes
involving transverse gauge bosons suffer from much larger backgrounds
which would obscure the effects of the symmetry-breaking sector.

Many terms in our effective Lagrangian include pieces that
correspond to four-point vertices, but only two are relevant here.
Since we care only about the
four-point scattering of longitudinal weak bosons, we can work in terms
of the
ungauged Lagrangian \lochiralun\ and \lagfour.  This eliminates
the $L_9$ and $L_{10}$ terms, for example.  Furthermore, the
contributions of the leading-order Lagrangian \lochiralun\
will, by the low-energy theorems, be identical in any
symmetry-breaking sector with a given
symmetry structure.  This leaves us with the order $p^4$
Lagrangian \lagfour ; as we assuming an $SU(2)_L\times
SU(2)_R/SU(2)_{L+R}$ global symmetry-breaking pattern, only the terms
proportional to $L_1$ and $L_2$ are present.

There are several different physical scattering processes encompassed
in the expression $V_L V_L \to V_L V_L$.  Since we are assuming that
our theory possesses an unbroken custodial $SU(2)_{L+R}$ symmetry, the
scattering amplitudes for the different proceses are related to one
another by crossing and $SU(2)_{L+R}$ symmetries.  More precisely, if the
amplitude for the process $W^+_L W^-_L \to Z_L Z_L$ is given by
\eqn\wpmdef{
{\cal M}(W^+_L W^-_L \to Z_L Z_L) \equiv A(s,t,u)
{}~~~,
}
where s,t,u are the usual Mandelstaam kinematic variables, then
the amplitudes for the other $V_L V_L \to V_L V_L$ processes are
\eqn\wpmdeff{
\eqalign{
{\cal M}(W^+_L W^-_L \to W^+_L W^-_L) = &\ A(s,t,u) + A(t,s,u)\cr
{\cal M}(Z_L Z_L \to Z_L Z_L) = &\ A(s,t,u) + A(t,s,u) + A(t,s,u)\cr
{\cal M}(W^\pm_L Z_L \to W^\pm_L Z_L) = &\ A(t,s,u)\cr
{\cal M}(W^\pm_L W^\pm_L \to W^\pm_L W^\pm_L) = &\ A(t,s,u) + A(u,t,s)
{}~~~.
}
}
The tree-level contribution to $A(s,t,u)$ from the low-energy theorem is
\eqn\letmm{
A(s,t,u)_{L.E.T.} = {s\over{f^2}}
{}~~~.
}
The tree-level contribution at order $p^4$ from \lagfour\ is \dvb
\eqn\letmmfour{
A(s,t,u)_4 = {4\over f^4}\left( 2 L_1 s^2 + L_2 (t^2 + u^2)\right)
{}~~~.
}
The form and symmetries of the amplitudes for the 2-body scattering of
Nambu-Goldstone bosons arising from the effective Lagrangian \lagfour\
have been discussed at length in \cdg\gasser .

The amplitudes we have written down for $V_L V_L \to V_L V_L$
scattering make it clear that production of all of the different $V_L
V_L$ final states will be affected by the order $p^4$ effective
Lagrangian coefficients $L_1$ and $L_2$.  However, some final states
lend themselves more readily to the study of four-point vertices than
others do.  The $W_L^+ W_L^-$ and $W^\pm_L Z_L$ final states are
produced mostly through $f\bar f$ annihilation rather than weak boson
re-scattering; therefore production of these states is more sensitive
to alteration of the 3-gauge-boson vertex by the $L_9$ terms of
\chiral\ than to alteration of the 4-point vertex by $L_1$ or $L_2$
\fls\bdv.  The $Z_L Z_L$ final state cannot be produced through a
3-weak-boson vertex and therefore lacks the large $f\bar f$
annihilation background; however there are backgrounds from continuum
$Z Z$ production and (at hadron colliders) from gluon fusion through a
top quark loop \golddd .  The $W^\pm_L W^\pm_L$ final state has the
distinct advantage of being free from order $\alpha^2$ continuum
backgrounds; the largest backgrounds are from $t\bar t$ production and
decay and from a mixed electroweak-strong process in which a gluon is
exchanged between two initial-state quarks which then each radiate a
weak boson.  The lower background rates in this channel should allow
the observation of signal events at relatively low invariant mass;
since the $WW$ distribution functions fall with increasing invariant
mass, this increases the signal-to-background ratio.  Indeed, the
$W^\pm_L W^\pm_L$ channel appears to be most sensitive to $L_1$ and
$L_2$ once backgrounds, branching fractions and cuts are taken into
account \goldch\bdv\golddd\canon.

A good deal of effort has been directed at estimating the ability of
proposed high-energy colliders to constrain $L_1$ and
$L_2$\foot{Current LEP data have been shown to place indirect bounds
of order 20 on $L_1$ and $L_2$\lepcalcdv.  These bounds are based on
loop-level calculations assuming that no large tree-level contribution
causes a significant cancellation of the effect.}. It has been found
that an NLC can probe the coefficients $L_1$ and $L_2$ down to the
level of 1-5 \fbj.  The LHC is projected to do even better --
measuring them to within their natural size of order 1
\golddd\bdv\wudka .


\newsec{Beyond the Effective Lagrangian}

The effective Lagrangians discussed in previous sections can never provide
anything more than a low energy description of symmetry breaking physics.
Since they are non-renormalizable, effective Lagrangians cannot be extended to
arbitrarily high energies.  Ultimately one wants to know the true
structure of the strongly interacting theory.

Consider the interactions of the ordinary hadrons.  The effective
Lagrangian for the low-energy states in QCD describes only the
scattering of pions near zero momentum.  At energies above a few GeV one
may use perturbative QCD to describe features of the physics such as the
rate of multijet events.  In a sense, it is most difficult to describe
the range of energies between approximately 1 and 10 GeV.  This is the
region that contains bound state resonances such as the $\rho(770)$ and
the baryons.  The techniques for describing this region are nowhere near
as simple and beautiful as those that work for either low or high
energies.

In this section we wade into the bog of intermediate energy.  We will go
beyond the dynamics of the longitudinal gauge bosons, to describe what happens
when other resonances appear.  This discussion is speculative, because
even in QCD the understanding of this physics is difficult.  No one really
knows how it will look in theories that are significantly different.

\subsec{QCD-like Theories}

In a technicolor theory with one doublet of techniquarks and three
technicolors, the spectrum should be the same as that of QCD, but scaled
up by a factor of $v/f_\pi$ \dimop.  Furthermore, all of the
interactions of these techniparticles would be expected to mirror those
of the corresponding particles in QCD.  For example, the ratio of the
mass to the width of the technirho should be the same as that of the
$\rho(770)$.  This scenario is the simplest technicolor model (though it
is probably disallowed by experiment).

Gauging the lowest-order (two-derivative) effective Lagrangian for this
model is not very interesting.  As we saw in the last section, in
unitary gauge the symmetry breaking sector is nothing more than a mass
term for the gauge bosons, with no hint of the structure of the higher
energy theory.  It is therefore imperative to find some way of going
beyond the lowest-order Lagrangian.

Section 2 discussed the most straightforward approach: include the
four-derivative terms of the effective Lagrangian.  As shown there, this
yields a description of the three- and four-gauge-boson vertices.
However, this method is not really adequate for the energy regime in
which a strongly interacting theory can be expected to become
distinctive: only in the region above 1 TeV will one expect that the
$\pi^a$ scattering amplitudes become strong, leading to the formation of
resonances.

Consider the analogous situation in the ordinary strong interactions.
At present, there is no ideal way of parametrizing the energy region
in which the bound-state resonances occur.  The techniques currently
used to describe the plethora of resonances, such as the
non-relativistic quark model, are {\it ad hoc} and not based on either
chiral symmetry or QCD. In a QCD-like technicolor theory it is
essential to preserve the chiral symmetry, because that symmetry is
gauged.  What is frequently done is to include the lightest resonances
into the effective Lagrangian.  So long as the dynamics are QCD-like,
the lightest particles are expected to be the vector resonances, the
technirho and techniomega, which are the analogues of the $\rho(770)$
and $\omega(783)$.  One hopes that in this way at least some of the
intermediate-energy region can be described.

We now discuss two equivalent methods for including  the vector
resonances into the effective Lagrangian \CCWZ. The
simplest way to include the $\rho(770)$ into the effective Lagrangian for
the strong interactions is as a ``matter'' field, an object that
transforms homogeneously under chiral rotations.
In order to include matter fields, one proceeds in two steps.  First, a field
$\xi$ is defined by
\eqn\xidef{
\xi^2 = \Sigma
{}~~~.
}
This field does not transform linearly under $SU(2)_L \times SU(2)_R$
rotations.  Instead one must define the matrix $U$ by
\eqn\xixform{
\xi \to L \xi U^\dagger = U \xi R^\dagger
{}~~~.
}
Note that $U$ will depend on $\xi$ and hence on spacetime.  The $\rho$
field can now be included.  It written as a matrix
\eqn\RRdef{
\rho^\mu = {1 \over 2} \rho^\mu_i \sigma_i
{}~~~,
}
where $\sigma_i$ are the Pauli matrices and $\rho^\mu_i$ are three real fields.
This field is taken to transform as
\eqn\RRxform{
\rho^\mu \to U \rho^\mu U^\dagger
{}~~~.
}
The kinetic energy term for $\rho$ is then
\eqn\LKEone{
{\cal L}_{KE} = -{1 \over 2} \tr (d^\mu \rho^\nu - d^\nu \rho^\mu)^2
{}~~~,
}
where $d^\mu$ is a chirally covariant derivative, defined by
\eqn\covone{
d^\mu \rho^\nu \equiv
\partial^\mu \rho^\nu + i V^\mu \rho^\nu - i \rho^\nu V^\mu
{}~~~,
}
and
\eqn\Vmu{
V^\mu = -{i \over 2} (\xi^\dagger \partial^\mu \xi + \xi \partial^\mu
\xi^\dagger)
{}~~~.
}
This is a chirally covariant derivative, in the sense that
\eqn\covd{
d^\mu \rho^\nu \to d^\mu (U \rho^\nu U^\dagger) =
U (d^\mu \rho^\nu) U^\dagger
{}~~~.
}
The mass term for the $\rho$ is
\eqn\Lmone{
{\cal L}_m = m_\rho^2 \tr \rho^\mu \rho_\mu
{}~~~.
}
The chirally covariant derivative can act on the fields of other particles
that are included in the effective Lagrangian.  For example, if the nucleon
doublet $N$ is included, it transforms as $N \to U N$.  Its kinetic energy
and mass terms are
\eqn\Lbar{
{\cal L}_B = {\bar N} (i {\slash d} + m_N) N
{}~~~.
}
One then proceeds to add all possible chirally invariant terms to the
Lagrangian.  For example, the term that couples the $\rho$ to the nucleon is
just
\eqn\Lcoupone{
{\cal L}_{\rho N N} = g_\rho {\bar N} {\slash \rho} N
{}~~~.
}
In this formulation, it is a mystery why the $\rho$ appears to dominate the
vector current form factor of the nucleon, and why its couplings appear to be
universal.

Frequently, a different approach is used: the $\rho(770)$ is included as a
gauge particle of a broken ``hidden local symmetry''\foot{When this method of
including the vector mesons is used in the electroweak theory, it is known as
the BESS model\BESS.}.  The goal is to give some explanation of the
universality
of the $\rho$ couplings to other particles. One can
show the equivalence of the two approaches by defining
\eqn\RRnew{
\rho_{new} = \rho + g_\rho V^\mu
{}~~~,
}
where here $g_\rho$ acts as a gauge coupling constant.  The chirally covariant
derivative that acts on other fields in the Lagrangian is
\eqn\covtwo{
d_{new}^\mu = \partial^\mu + i g_\rho \rho_{new}^\mu
{}~~~.
}
One may as usual define a ``field strength'' tensor
\eqn\Fmn{
F^{\mu\nu} = {1 \over i} [d_{new}^\mu, d_{new}^\nu]
{}~~~,
}
and then the kinetic energy term of the $\rho$ meson is
\eqn\LKEtwo{
{\cal L}_{KE} = - {1 \over 2} \tr F^{\mu\nu} F_{\mu\nu}
{}~~~,
}
and the mass term of the $\rho$ field is
\eqn\Lmtwo{
{\cal L}_m = m_\rho \tr (\rho_{new}^\mu - g_\rho V^\mu)^2
{}~~~.
}
When an effective Lagrangian is built out of $d_{new}$, the $\rho$ has
universal couplings to the nucleons and other particles.
However, any chirally invariant
term that can be built in the ``matter'' formulation of the previous paragraph
will continue to appear in this approach. In other words, it is still
possible to add additional $\rho$--nucleon couplings by adding additional
operators, such as
\eqn\Lcouptwo{
{\cal L}_{\rho NN} = h {\bar N} ({\bf \slash \rho}_{new} - g_\rho {\slash V}) N
{}~~~.
}
The mystery of the universality of the $\rho$ couplings persists in
this formulation
of the theory, but now it takes a different form.  Now we need to know why the
new coupling $h$ is so much smaller than one expects.

While it is true that every vertex in the ``matter field $\rho$'' Lagrangian
can be written in the ``gauge field $\rho$'' Lagrangian, the power counting of
the operators is somewhat different in the two approaches.  In the latter
approach, $m_\rho$ gets renormalized only by terms proportional to powers of
$g_\rho$.  Only in the ``gauge field $\rho$'' method can one understand a
light, weakly coupled vector.

On the other hand, it is not really strictly valid to include the $\rho(770)$
in the effective Lagrangian description of QCD.  The effective Lagrangian
breaks down at a scale near $\Lambda$.  If $\Lambda$ were much
greater than 770 MeV, then one would expect to see a big hierarchy between the
$\rho(770)$ and other physics.  This does not appear to be the case; once 770
MeV is reached the resonances come thick and fast.  Another way of saying this
is that when the $\rho$ is included as a gauge field, the coupling $g_\rho$ is
of order $4\pi$.  By analogy, it is likely to be invalid in a strict sense to
include the technirho in the effective Lagrangian for a technicolor sector,
because neither formulation has a valid procedure for the inclusion of
technirho loops or higher-derivative multipion operators.

However, for the simpler purposes of calculating event rates, inclusion
of the technirho into the effective Lagrangian and working at tree level
is actually quite useful.  It does yield a qualitatively reasonable,
gauge invariant amplitude for gauge boson scattering.  There have been
numerous papers that have looked at the possibility of seeing the
low-lying hadronic resonances at colliders\chan\equivb\tehlq\coll, but
many of these papers discuss observability at 17 TeV or 40 TeV machines.
At present, more work is needed to determine exactly where the window of
discovery is at the 14 TeV LHC.

\subsec{Limits of the Effective Lagrangian}

In this paper we have mentioned examples of technicolor models with various
numbers of flavors.  In such models the symmetry breaking pattern is
$SU(N_f)_L \times SU(N_f)_R \to SU(N_f)_{L+R}$.  For example, $N_f=2$ in the
simplest technicolor model and $N_f=8$ in the one-family model.  We
begin by deriving some formulas implementing the extension to $SU(N_f)$
of the custodial symmetry.  We will use these results to explore the scale
$\Lambda$ at which the effective Lagrangian breaks down, and to argue
that the number of flavors directly influences the mass of resonances,
such as the technirho \cdg.

As stated in the introduction, the chiral symmetries of technicolor are
generally only approximate symmetries. In addition to the three
``eaten,'' exact Nambu-Goldstone bosons, there are often additional
pseudo-Goldstone bosons.  If the mass of a typical pseudo-Goldstone
boson is $m$, the effective Lagrangian is also an expansion in $m^2$.
For simplicity, we consider a chiral symmetry breaking interaction that
does not break the conserved $SU(N_f)_{L+R}$ vector symmetry.  Such a chiral
symmetry breaking term gives the same mass to all Nambu--Goldstone bosons.

Consider the scattering process $\pi^a \pi^b \to \pi^c \pi^d$.  The
amplitude for such a process may be decomposed into irreducible
representations of the unbroken $SU(N_f)_{L+R}$.  Since the $\pi$'s are in
the adjoint representation of this symmetry, one needs to know the
representation content of adjoint $\otimes$ adjoint.  For $N_f=2$ there
are three representations, corresponding to the isospin 0, 1, and 2
channels. For $N_f=3$, the representations are the familiar $\bf 1, 8_a,
8_s, 27, 10,$ and $\bf \overline{10}$ in ${\bf 8} \otimes {\bf 8}$.  For
$N_f > 3$, there are always seven representations.  Of greatest interest
for our purposes will be the singlet representation, in which the
incoming $\pi$'s have the same flavor: $a=b$.

We can construct the most general amplitude for $\pi\pi$ scattering
consistent with Bose symmetry, crossing invariance, and $SU(N_f)_{L+R}$
conservation. If we define $d^{abc}$ and $f^{abc}$ by
\eqn\fdef{
f^{abc} = -2i \tr [T^a,T^b] T^c \hbox{\hskip .2in and \hskip .2in}
d^{abc} =  2  \tr \{T^a,T^b\} T^c
{}~,}
then the most general amplitude is
\eqn\smatrix{
\eqalign{
a(s,t,u)^{a,b;c,d} =&
   \delta^{ab}\delta^{cd} A(s,t,u)
 + \delta^{ac}\delta^{bd} A(t,s,u)
 + \delta^{ad}\delta^{bc} A(u,t,s)\cr
&+ d^{abe} d^{cde} B(s,t,u)
 + d^{ace} d^{bde} B(t,s,u)
 + d^{ade} d^{bce} B(u,t,s)~~~,
}
}
where $s$, $t$, and $u$ are the Mandelstam variables and $A$ and $B$ are
unknown functions.  Bose symmetry implies that the functions $A$ and $B$ must
be symmetric under the exchange of their second and third arguments. From
this amplitude we may derive the scattering amplitude of, for example,
the singlet representation:
\eqn\aAB{
\eqalign{
a_0(s,t,u) =&  (N_f^2-1)A(s,t,u)+A(t,s,u)+A(u,t,s) \cr
	    &+ {(N_f^2-4)\over N_f}(B(t,s,u)+B(u,t,s))
{}~~~.
}
}
We may, as usual, project this amplitude onto its various (even) orbital
angular momentum components.  The functions $A$ and $B$ will be
such that all these partial wave amplitudes obey the usual unitarity
relations.

One may use the two-derivative effective Lagrangian to compute the invariant
functions $A$ and $B$: $A(s,t,u)=(2/N_f)(s-m^2)/f^2$, and
$B(s,t,u)=(s-m^2)/f^2$.  The isosinglet spin-zero scattering amplitude is
therefore \cs:
\eqn\scatt{
a_{00} = {N_f s \over 32 \pi f^2} - {m^2 \over 16 \pi N_f f^2}
{}~~~.
}
It will be important below to note that this scattering amplitude is enhanced
by a factor of $N_f$.

We have seen in previous sections that the most general chirally
invariant Lagrangian can be written as an expansion in powers of
derivatives.  Additional terms with more derivatives are suppressed by
powers of the momentum scale that we have denoted $\Lambda$.  The
effective Lagrangian is an expansion in $p^2/\Lambda^2$ (and
$m^2/\Lambda^2$).

At energies near or above $\Lambda$, all terms in the expansion
contribute and the effective Lagrangian becomes useless.
The amplitude $a_{00}$ calculated at tree level is real, and (for small
$m^2$) exceeds 1 when $\sqrt{s} > 4 \pi f / \sqrt{N_f}$.  A physical
scattering amplitude must lie on or inside the Argand circle, but the
point $a_{00}=1$ is far outside.  At these energies, therefore, loop
corrections and higher order terms in the effective Lagrangian must make
as large a contribution as the two-derivative term, and the calculation
using just the lowest order effective Lagrangian ceases to be
useful. This suggests that $\Lambda$ is less than or of order $4 \pi
f/\sqrt{N_f}$, as was emphasized in \SS \foot{Note that in section 2,
$N_f = 2$ and the factor of $1/\sqrt{N_f}$ was neglected.}.

An alternative approach which puts the same limit on $\Lambda$ is based
on an estimate of the size of loop corrections \wein.  Since the theory is not
renormalizable, the terms of order $p^4$ are required as counterterms to loops
involving the lowest order interactions.  In calculating the scattering
amplitude to order $p^4$, one must consider tree-level diagrams with
interactions coming from operators of fourth order in momenta, and one-loop
diagrams using the two-derivative terms in the effective Lagrangian.
Similarly, the two-loop calculation using the lowest order effective
Lagrangian will require counterterms of order $p^6$, etc.

In doing the one-loop calculation, it is unnatural to assume that the
contribution from the loop diagrams is much larger than that from the
tree-level four-derivative operators, since such a statement could only
be true for a particular choice of renormalization scale\foot{The
explicit calculation \cdg\ of the one loop corrections to the tree-level
functions $A$ and $B$ shows that the results are factors of order $N_f
s/(16 \pi^2 f^2)$ or $N_f m^2/(16 \pi^2 f^2)$.}.  Therefore it is
inconsistent to assume that the corrections to $\pi\pi$ scattering of
order $p^4$ are less than or of order $(\sqrt{N_f} p / 4\pi f)^2$ where
$p$ is a typical momentum in the process -- meaning that it is
unrealistic to assume that the coefficients of the higher order
four-derivative terms in the effective Lagrangian are smaller than about
$N_f/16\pi^2 f^2$.

It is possible to show that this pattern persists to all orders: with each
additional loop, the corrections are a factor of order $(\sqrt{N_f} p / 4\pi
f)^2$ times the previous correction.  Again, this implies that at any order in
the momentum expansion, the mass scale $\Lambda$ suppressing the
higher derivative terms cannot be much larger than $4 \pi f / \sqrt{N_f}$.

\subsec{Implications for New Physics}

Some interesting questions arise at this point.  We have argued that the
momentum expansion breaks down at or before $\Lambda$, but what actually
happens to the amplitudes as $s$ increases beyond this value?  What is the
significance of $\Lambda$?  The amplitudes for the partial waves other
than $a_{00}$ are all below their unitarity limits when $\sqrt{s} = 4
\pi f / \sqrt{N_f}$.  Is it possible that, like $\Lambda_{QCD}$, $\Lambda$
is a purely calculational artifact corresponding to no particular
physical structure?\foot{Ref \AT\ argues for this interpretation.  In
the absence of any well-motivated way to calculate these field theories,
any such argument, including the one presented in this paper, is fairly
speculative.  However, in certain toy models the argument presented here
can be made rigorous \dugol.}

In the effective Lagrangian the multiderivative terms contribute an
arbitrary polynomial in $s$, $t$, and $u$ to the scattering amplitude:
\eqn\series{
\sum_k a_k {p^2 \over f^2} \left({p^2 \over \Lambda^2}\right)^{k-1}
}
where all the $a_i$ are numbers of order 1.  When does such a series fail to
converge?  Since the $a_k$ are of order one, the radius of convergence is
$\Lambda$.  Because the series diverges at energies higher than
$\Lambda$, the momentum expansion (to any, arbitrarily high, finite
order) cannot give a good approximation to the scattering amplitude at
energies beyond $\Lambda$.

It is plausible that the effective Lagrangian can accurately match the
scattering amplitude out to the first non-analytic structure representing new
physics.  This is because the $S$ matrix is an analytic function of momenta,
except at isolated points where intermediate states go on shell.  All the
non-analytic structures in the scattering amplitudes corresponding to
multipion states are correctly included at some order in the momentum
expansion by the pion loop calculations - what is not properly included are
effects of other states.  For example, in QCD there is a pole in the $S$
matrix at the $\rho(770)$ mass.  Above $m_\rho$, the effects of a term like
$1/(p^2-m_\rho^2)$ in the $S$-matrix this can never be reproduced as a finite
power series in positive powers of $p$.  The series has to be resummed in some
way.

If this argument is correct, then it follows that $\Lambda$ is precisely
the mass of the lightest non-analytic structure in the $S$-matrix.  The
conclusion is that new physics is lighter than a scale of order $4 \pi f /
\sqrt{N_f}$.

The implications for technicolor and strongly interacting field theories
in general may be substantial.  In the case of technicolor, it may be
that the new physics that comes in at this low scale is the technirho.
If this is true then the technirho mass suggested by the simple scaling
argument may be a significant overestimate.  In this case, if the vector
dominance relations continue to hold, then the simplest estimates of
oblique radiative corrections in technicolor models, such as those
in \chilg\ and \pestak, may be rather unreliable \reconsid.  If
technicolor somehow manages to evade the problems from radiative
corrections, the lightness of the vector bosons may make them
interesting for future colliders \eephenom.

Even if non-perturbative physics has nothing to do with electroweak symmetry
breaking, the arguments of this section may be of interest for ordinary QCD.
The idea that the masses of the resonances depend on the number of
light flavors may seem counterintuitive.  If the charm, strange,
top, and bottom quarks were as light as the up and down, the ratio $m_\rho /
f_\pi$ would be substantially altered.  However, in a non-relativistic quark
model neither the $\rho$ nor $\pi$ contains anything other than the first
generation quarks.  Understanding the effects discussed in this section may
have something to do with understanding the difficulties \Qprob\ with quenched
chiral perturbation theory \Quenched.

\subsec{A ``Hidden'' Symmetry Breaking Sector}

As we have seen, the most direct probe of the symmetry breaking sector
is the scattering of longitudinal $W$ and $Z$ bosons.  That is because
at energies large compared to their mass, the longitudinal components of
these particles are (essentially) the eaten Nambu--Goldstone Bosons of
$SU(2)_W \times U(1)_Y$ symmetry breaking \equivb.  Section 1
discussed various possibilities for the new physics that enters at the
scale $\Lambda$: in the weakly coupled one-doublet Higgs model, it was
the light and narrow Higgs boson; in minimal technicolor, the exchange
of the technirho and other particles unitarizes gauge boson scattering.

It is frequently assumed that these two types of behavior for elastic $W$ and
$Z$ scattering are generic (see, for example \chan,\canon). If the symmetry
breaking sector is weakly coupled, the growth of the $W_L W_L$ scattering
amplitudes is cut off by narrow resonances (like a light Higgs boson) at a
mass scale well below a TeV. For strongly coupled theories, it is assumed that
the amplitudes saturate unitarity and that there are broad resonances in the
TeV region where the strong interaction sets in.

There is another possibility: if the electroweak symmetry breaking
sector has a large number of particles, the {\it elastic} $W$ and $Z$
scattering amplitudes can be small and structureless, i.e. lacking any
discernible resonances.  Nonetheless, the theory can be strongly
interacting and the {\it total} $W$ and $Z$ cross sections large: most
of the cross section is for the production of particles other than the
$W$ or $Z$. In such a model, termed a ``Hidden Symmetry Breaking
Sector'' \hidden, discovering the electroweak symmetry breaking physics
depends on the observation the other particles and the ability to
associate them with symmetry breaking.  Physicists should keep an open
mind about the experimental signatures of the electroweak symmetry
breaking sector because discovery of electroweak symmetry breaking may
not rely solely on two-gauge-boson final states.

This scenario may be illustrated by considering a toy model of the
electroweak symmetry breaking sector based on an $O(N)$ linear sigma
model.  This model is particularly interesting since it can be solved
(even for strong coupling) in the limit of large $N$ \Coleman.  One
constructs a model with both exact Nambu--Goldstone bosons (which will
represent the longitudinal components of the $W$ and $Z$) and
pseudo-Goldstone bosons. To this end let $N=j+n$ and consider a model
with $j$- and $n$-component real scalar fields.  One can construct a
theory that has an approximate $O(j+n)$ symmetry which is broken softly
but explicitly to $O(j) \times O(n)$.  A vacuum expectation value,
breaks the $O(j)$ symmetry to $O(j-1)$, and the theory has $j-1$
massless Nambu--Goldstone bosons and one massive Higgs boson. The $O(n)$
symmetry is unbroken, and there are $n$ degenerate massive
pseudo-Goldstone bosons, which we refer to as $\psi$s.  It is possible
to solve this model in the limit that $j,n \to \infty$ with $j/n$ held
fixed\foot{For the complete details of the construction and solution of
this model, see \hidden\ and \scalars.}.

The scalar sector of the standard one-doublet Higgs model has a global $O(4)
\approx SU(2) \times SU(2)$ symmetry, where the 4 of $O(4)$ transforms as one
complex scalar doublet of the $SU(2)_W\times U(1)_Y$ electroweak gauge
interactions. This symmetry is enlarged in the $O(N)$ model: the spin-0
weak isosinglet scattering amplitude of longitudinal gauge bosons is
modeled by the spin-0 $O(j)$ singlet scattering of the Nambu--Goldstone
bosons in the $O(j+n)$ model solved in the large $j$ and $n$ limit.  Of
course, $j=4$ is not particularly large. Nonetheless, the resulting
model will have all of the qualitative features needed, and the
Nambu--Goldstone boson scattering amplitudes will be unitary (to the
appropriate order in $1/j$ and $1/n$).  Thus this theory can be used to
investigate the scattering of Nambu--Goldstone bosons at moderate to
strong coupling \Einhorn.  Since they are mostly produced via their
strong interactions, the electroweak quantum numbers of the
pseudo-Goldstone bosons can be anything; here we assume that the
pseudo-Goldstone bosons are $SU(3)$ color singlets\foot{Gauge boson pair
production in models with colored pseudo-Goldstone bosons is discussed
in detail in \gscat.}.

Nambu--Goldstone boson scattering in an $O(N) \to O(N-1)$ model is in
some ways similar to that in the $SU(N_f) \times SU(N_f)$ model considered
above.  For example, the amplitude $a^{ij;kl}(s,t,u)$ for the process
$\pi^i \pi^j \to \pi^k \pi^l$ is
\eqn\scatamp{
a^{ij;kl}(s,t,u) = A(s;M)\delta^{ij}\delta^{kl} +
A(t;M)\delta^{ik}\delta^{jl} +
A(u;M)\delta^{il}\delta^{jk}
{}~~~,
}
where $A(s;M)$ is some function.  This $O(N)$ theory is soluble to leading
order in $1/N$, so $A$ may in fact be computed to this order without any
assumptions.  In the equation above $M$ is a parmeter with dimensions of mass
that specifies the strength of the self-coupling of the symmetry breaking
sector.  It is essentially a cutoff, and so the smaller $M$ is, the stronger
the self-coupling.  The isospin-zero amplitude spin-zero is therefore
calculable to order $1/N$ too; it is
\eqn\scatsing{
a_{00}(s) = {j A(s;M) \over 32 \pi}
{}~~~.
}

Plotted in \hidfig\ is the absolute value of $a_{00}$ {\it vs.} the
center-of-mass energy for different values of $M$.  We have set $j = 4$, as
always.  We have also set the number of pseudo-Goldstone bosons, $n$, to 32.
The pseudogoldstone bosons have a mass $m_\psi =$ 125 GeV.  The curves plotted
correspond to approximately $8M / m_\psi$ = 10000, 600, 200, 100, and 60.  For
the weakly coupled theory, for example the 10000 curve, there is a light Higgs
boson which decays to $\pi$'s.  When the Higgs boson is light, its width is
more or less unaffected by the heavy $\psi$'s, and thus its properties are
identical to those of the Higgs boson of similar mass in the $O(j)$ model
\Einhorn.  As the Higgs resonance gets closer to the two $\psi$
threshold, it gets relatively narrower than it would have been were the
$\psi$'s absent.  As the theory becomes more strongly coupled still, the
resonance gets heavier and broader. Eventually, for small enough $M$, the
imaginary part of the location of the pole is so great that there is no
discernible resonance in $a_{00}$.

When the Higgs resonance is heavier than twice $m_\psi$, it no longer decays
exclusively to $\pi$'s, and thus the absolute value of the amplitude for
elastic $\pi\pi$ scattering never gets anywhere near 1.  Probability is
leaking out of this channel into that for the production of pairs of $\psi$'s.
For comparison, the dashed line shows the scattering amplitude in the limit
$m_\psi \to \infty$ with $M$ adjusted to produce a Higgs resonance at
approximately 500 GeV.

In the gauged model, the $\pi$'s are eaten by the gauge bosons, and become
their longitudinal components.  Therefore, $\pi\pi$ final states correspond
to two-gauge-boson events.  In this toy model the Higgs resonance may be light
but so broad that at no energy is the number of $WW$ or $ZZ$ events large;
discovering the Higgs boson depends on its observation in the two $\psi\psi$
channel.  Depending on how the $\psi$'s decay, this may be easy or hard.
Nonetheless, it is clear that an experiment looking for electroweak symmetry
breaking may not be able to rely exclusively on the two-gauge-boson events.
Parton level computations have indicated \hphenom\ that it is probably not
possible to detect this symmetry breaking sector at the proposed LHC by
examining the gauge-boson-pair modes exclusively\foot{This claim is
disputed in \kny.  There it was shown that the numbers of final state gauge
boson pairs from gauge boson scattering is roughly independent of $N$ if
$\sqrt{N} M$ is held fixed. This is because as $N$ increases for fixed
$\sqrt{N} M$, $M$ and the mass and width of the Higgs boson decrease like
$1/\sqrt{N}$. The increased production of Higgs bosons due to their smaller
mass is approximately cancelled by the Higgs boson's smaller branching ratio
into $W$s and $Z$s.  The number of signal events, therefore, is approximately
independent of $N$ and is the same as the number which would be present in the
standard model.  However, the {\it background} rate is much larger, because
the Higgs boson is much lighter, and this renders the signal unobservable.}.

This section has shown that it is possible for the $W$ and $Z$ scattering
amplitudes to be small and structureless: if the symmetry breaking sector
contains a large number of particles in addition to the longitudinal gauge
bosons, there may be light but very broad ``resonances''.  The ability to
discover the electroweak symmetry breaking sector depends on the observability
of technipions other than the longitudinal gauge bosons \tehlq.
Associating the technipions of this model with electroweak symmetry breaking
will be crucial.

\newsec{Conclusions}

In this review we have discussed theories in which the $W$ and $Z$
interactions become strong at an energy scale of order a TeV or less.
We began with a survey of the range of theories which have been
constructed to explain electroweak symmetry breaking. We argued that
the putative triviality of theories with fundamental scalar particles
implies that {\it any} theory with a large hierarchy between the weak
scale and the scale of the dynamics responsible for producing
electroweak symmetry breaking must be {\it weakly interacting}.  This
implies that if the $W$ and $Z$ interactions are strong at energies of
order a TeV or less, the physics responsible for electroweak symmetry
must become apparent at {\it the same energy scale}.

We then reviewed the use of the effective Lagrangian to
describe the physics of any strongly-interacting symmetry breaking
sector at energies lower than the mass of the lightest resonance.
Limits on the values of low-energy parameters ({\it e.g.} $S$ or
$L_{10}$, and $T$) provide the most significant constraints on the
strongly-interacting symmetry breaking sector.

In order to discover the physics of the symmetry breaking sector it
will be necessary to probe physics at energy scales of order a TeV.
In a strongly-interacting symmetry breaking sector we expect that a
plethora of new resonances will appear at these energies to cut off
the growth of the $W_L$ and $Z_L$ scattering amplitudes.  As we
discuss in the last section, the effective Lagrangian ceases to
be a useful description at an energy scale of order the mass of these
resonances. Further, we argued that, in order for the effective chiral
Lagrangian to be self-consistent, the mass scale of the resonances
must be lighter than or of order $4\pi f/\sqrt{N_f}$.

Finally, it is often assumed that if the $W$ and $Z$ interactions are
strong, there will always be large $W$ and $Z$ scattering cross section
at high energies.  We concluded with a description of the ``hidden''
symmetry breaking sector in which, although the $W_L$ and $Z_L$
interactions are strong, the {\it elastic} scattering amplitudes are
always small and structureless.

\centerline{\bf Acknowledgments}

RSC acknowledges the support of an Alfred P. Sloan Foundation
Fellowship, an NSF Presidential Young Investigator Award, and a DOE
Outstanding Junior Investigator Award.  MG acknowledges the support of
NSF National Young Investigator Award. EHS acknowledges the support of
an American Fellowship from the American Association of University
Women.

{\it This work was supported in part by the National Science
Foundation under grants PHY-9057173 and PHY-9218167, and by the
Department of Energy under grant DE-FG02-91ER40676.}

\listrefs

$$
\epsfxsize= 5.5truein \epsfbox{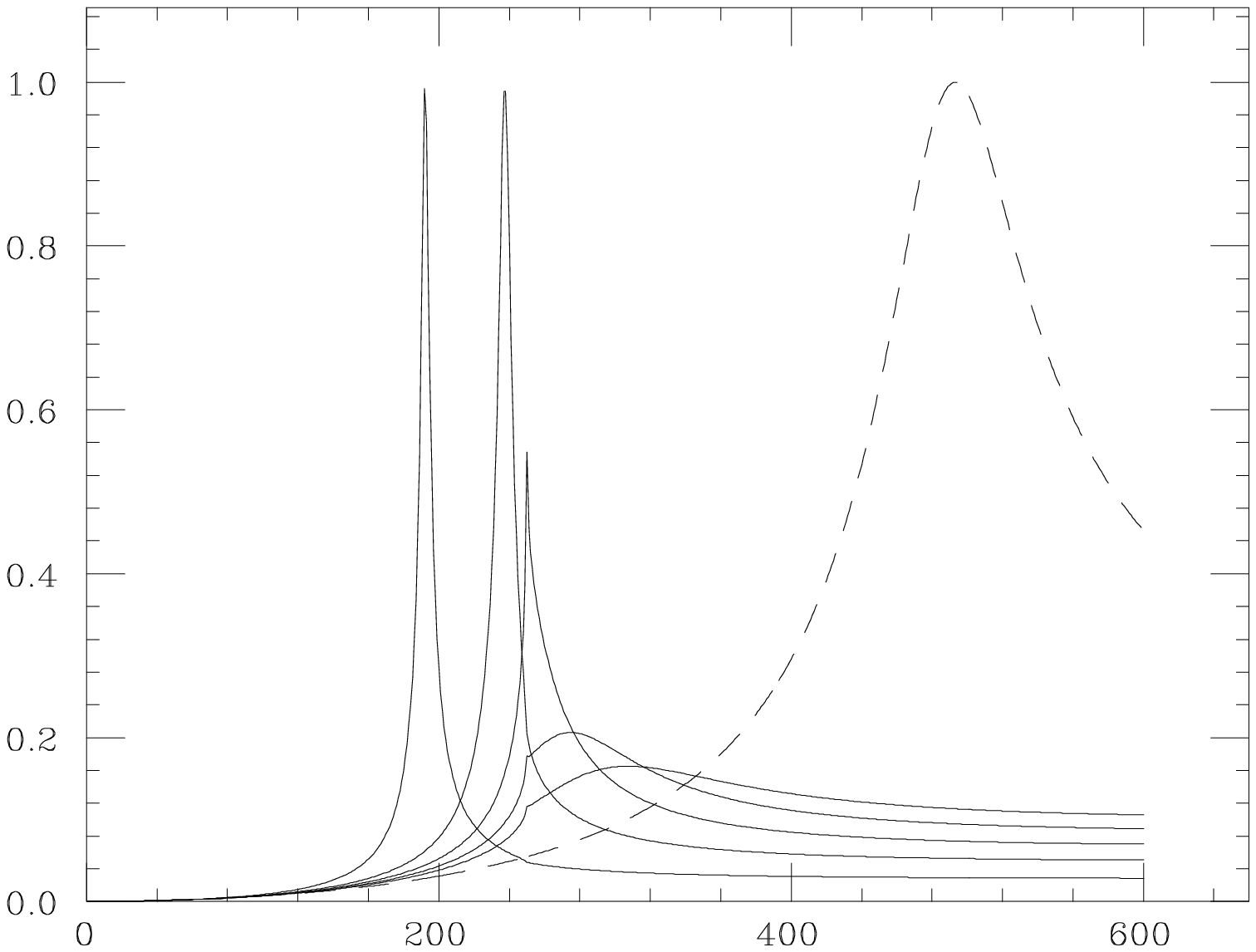}
$$

\listfigs

\bye